%% file: poll.tex
\documentclass[sigconf]{acmart}
\pdfoutput=1

\usepackage{booktabs} % For formal tables

\setcopyright{rightsretained}
\usepackage{amssymb,amsfonts,amsmath,amsthm,amscd,dsfont}%,dsfont,mathrsfs}
\usepackage{graphicx,float,epsfig}
\usepackage{wrapfig}
\usepackage{hyperref}
\usepackage{bbm}
\usepackage{authblk}
\usepackage{csquotes}
\usepackage{enumitem}
\usepackage{comment}
\usepackage{bbm,tikz}
\usepackage{subcaption}
\usepackage{tcolorbox}
\usepackage{prettyref,xspace,graphicx}
\usepackage{bbm,tikz}
\usepackage{authblk}
\usepackage[noend]{algpseudocode}

\newcommand{\scheme}{Barracuda}
\newcommand{\delay}{\Delta}

\newcommand{\diverge}{\to\infty}

\newcommand{\Expect}{\mathbb{E}}
\newcommand{\expect}[1]{\mathbb{E}\left[#1\right]}
\newcommand{\prob}[1]{\mathbb{P}\left[#1\right]}

\newcommand{\calD}{\mathcal{D}}

\newcommand{\pth}[1]{\left( #1 \right)}

\newtheorem{theorem}{Theorem}
\newtheorem{lemma}{Lemma}

\newtheorem{corollary}{Corollary}

\theoremstyle{definition}
\newtheorem{definition}{Definition}

\begin{document}
%\title{Information Balancing in Proof-of-Stake Blockchains:\\ 
%The Power of $\ell$-polling  }
\title{\scheme: The Power of $\ell$-polling in  Proof-of-Stake Blockchains  }
%\titlenote{Produces the permission block, and  copyright information}
%\subtitle{Extended Abstract}
%\subtitlenote{The full version of the author's guide is available as  \texttt{acmart.pdf} document}

%\authornote{Dr.~Trovato insisted his name be first.}
%\affiliation{%
%  \institution{Institute for Clarity in Documentation}
%  \streetaddress{P.O. Box 1212}

%\renewcommand{\shortauthors}{Giulia Fanti, Jiantao Jiao, Ashok Makkuva, Sewoong Oh, Ranvir Rana and Pramod Viswanath}

%\author{Giulia Fanti, Jiantao Jiao, Ashok Makkuva, Sewoong Oh, Ranvir Rana and Pramod Viswanath}
\affiliation{
  \institution{Giulia Fanti$^\dagger$, Jiantao Jiao$^\mathsection$, Ashok Makkuva$^\ddagger$, Sewoong Oh$^\ddagger$, Ranvir Rana$^\ddagger$ and Pramod Viswanath$^\ddagger$\\
  $^\dagger$Carnegie-Mellon University, $^\mathsection$University of California, Berkeley, $^\ddagger$ University of Illinois at Urbana-Champaign\\
  gfanti@andrew.cmu.edu, jiantao@eecs.berkeley.edu, \{makkuva2, swoh, rbrana2, pramodv\}@illinois.edu
  }
  }
%\email{gfanti@andrew.cmu.edu, jiantao@eecs.berkeley.edu, \{makkuva2, swoh, rbrana2, pramodv@illinois.edu}

\begin{abstract} 
A blockchain is a database of sequential events
that is maintained by a distributed group of nodes. 
A key consensus problem  in blockchains is that of determining the next block (data element) in the sequence.
Many blockchains address this by electing a new node to propose each new block.
The new block is (typically) appended to the tip of the proposer's local blockchain, and subsequently broadcast to the rest of the network.
%In blockchains, consistency  is ensured by sequentially elected proposers, 
%who check the validity of the current blockchain and 
%append at the end newly generated blocks of events. 
Without network delay (or adversarial behavior),
 this procedure would give a perfect chain, 
 since each proposer would have the same view of the blockchain. 
A major challenge in practice is {\em forking}.  
Due to network delays, 
a proposer may not yet have the most recent block,
and may therefore create a side chain that branches from the middle of the main chain.  
Forking reduces throughput, since only one a single main chain can survive, 
and all other blocks are discarded. 
%We propose a canonical network model to study the 
%trade-off between the network delay and the throughput. 
We propose a new P2P protocol for blockchains called  \scheme, in which each proposer, prior to proposing a block, polls $\ell$ other nodes for their local blocktree information. 
Under a stochastic network model, we prove that this lightweight 
primitive improves throughput as if the {\em entire} network were a factor of $\ell$ faster. %speed up of the entire network were improved by a factor of $\ell$. 
We provide guidelines on how to implement \scheme~ in practice, 
%with a specific emphasis on proof-of-stake blockchains, 
guaranteeing robustness against several real-world factors. 

%A blockchain is a database of sequential events, such as transactions of cryptocurrency,  
%that guarantees consistency. 
%Consistency is ensured by sequentially elected proposers, 
%who check the validity of the current blockchain and 
%append at the end newly generated blocks of events. 
%Ideally, when there is no network delay, 
% this process produces a perfect chain, 
% as every proposer has the same up-to-date view of the blockchain. 
%
%One of the major challenges in practice is {\em forking}.  
%Due to network delays, 
%a proposer might not have received a block. 
%This results in forking, 
%where a new block creates a side chain, branching out from the middle of the main blockchain.  
%This reduces the throughput of the system, as 
%eventually a single main chain will survive, 
%and all other blocks will be discarded. 
%%We propose a canonical network model to study the 
%%trade-off between the network delay and the throughput. 
%Our main result is the proposal of a new P2P protocol for blockchains that we call  $\ell$-polling. 
%Under a canonical stochastic network model, we prove that lightweight 
%polling primitives ameliorate the informational imbalance strongly: the resulting throughput is as 
%if  the speed up of the entire network were improved by a factor of $\ell$. 
%We provide guidelines on how to implement $\ell$-polling in practice with specific relevance to proof of stake 
%blockchains, guaranteeing robustness against several real-world factors. 

\end{abstract}

%<ccs2012>
%<concept>
%<concept_id>10003033.10003039.10003040</concept_id>
%<concept_desc>Networks~Network protocol design</concept_desc>
%<concept_significance>500</concept_significance>
%</concept>
%</ccs2012>
%\end{CCSXML}
%
%\ccsdesc[500]{Networks~Network protocol design}

\begin{CCSXML}
<ccs2012>
<concept>
<concept_id>10002950.10003648.10003671</concept_id>
<concept_desc>Mathematics of computing~Probabilistic algorithms</concept_desc>
<concept_significance>500</concept_significance>
</concept>
<concept>
<concept_id>10003033.10003039.10003051.10003052</concept_id>
<concept_desc>Networks~Peer-to-peer protocols</concept_desc>
<concept_significance>500</concept_significance>
</concept>
<concept>
<concept_id>10010520.10010575</concept_id>
<concept_desc>Computer systems organization~Dependable and fault-tolerant systems and networks</concept_desc>
<concept_significance>300</concept_significance>
</concept>
</ccs2012>
\end{CCSXML}

\ccsdesc[500]{Mathematics of computing~Probabilistic algorithms}
\ccsdesc[500]{Networks~Peer-to-peer protocols}
\ccsdesc[300]{Computer systems organization~Dependable and fault-tolerant systems and networks}

\keywords{Stochastic networks, blockchains }

\copyrightyear{2019} 
\acmYear{2019} 
\setcopyright{acmcopyright}
\acmConference[Mobihoc '19]{The Twentieth ACM International Symposium on Mobile Ad Hoc Networking and Computing}{July 2--5, 2019}{Catania, Italy}
\acmBooktitle{The Twentieth ACM International Symposium on Mobile Ad Hoc Networking and Computing (Mobihoc '19), July 2--5, 2019, Catania, Italy}
\acmPrice{15.00}
\acmDOI{10.1145/3323679.3326533}
\acmISBN{978-1-4503-6764-6/19/07}

\maketitle

\input{intro}
\input{sec2} % Random process of block chains 
\input{sec3} % Throughput analysis
\input{sec4} % Benefits of push-pull protocol
\input{sec5} % System & implementation issues 
\input{sec6} % Proofs 

\input{related}

\input{conclusion}

% \acknowledgement
\begin{acks}
We thank Sanjay Shakkottai for helpful discussions on the impact of polling on load balancing.
This work was supported by NSF grants CCF-1705007,   CCF-1617745,  and CNS-1718270, ARO grant W911NF-18-1-0332-(73198-NS), the Distributed Technologies Research Foundation, and Input Output Hong Kong. 
\end{acks}

% don't forget to acknowledge Sanjay Shakkottai!!

\bibliographystyle{ACM-Reference-Format}
\bibliography{blockchain}

\end{document}

%% file: intro.tex
% !TEX root =  poll.tex

\section{Introduction}
\label{sec:intro}

%goal is to form a chain, but due to randomness and network delay, we get a tree. 

%The word `blockchain' has two meanings. 
Blockchains are a sequential data structure in which each element  depends in a structured, predefined manner on every prior element. 
Most blockchains implement this property recursively by including in each data element a hash of the previous element.
This makes it easy to append an element to the end of a blockchain, but difficult to alter or insert elements in the middle of a blockchain, since every subsequent element must be modified to preserve validity. 
In parallel, the word `blockchain' has also come to mean the network and consensus algorithms that enable a distributed set of nodes to maintain such a data structure robustly and consistently. 

In practice, there are many obstacles to maintaining a distributed blockchain, including peer churn, adversarial behavior, and unreliable networks.
In this paper, we focus on the latter challenge and consider how to build efficient blockchains over unreliable networks. 
Although the research community is increasingly studying   peer-to-peer (P2P) networks in blockchain systems \cite{DW13,miller2015discovering,biryukov2014deanonymisation,heilman2015eclipse,falcon}, network effects are arguably the aspect of blockchains that have received the least attention thus far.
% , with a few exceptions \cite{falcon,bloxroute}.
In particular, we are interested in how the network affects blockchain performance metrics like latency and throughput for new data elements. 
To explain the problem, we start with a brief description of blockchain functionality. 

\bigskip
\noindent {\bf Blockchain Primer.} 
Blockchain systems are typically used to track sequential events, such as financial transactions in a cryptocurrency. 
A \emph{block} is simply a data structure that stores a batch of such events, along with a hash of the previous block contents.
The core problem in blockchain systems is determining (and agreeing on) the next block in the  data structure.
%To understand the network's role in this problem, consider a blockchain system in which 
Many leading cryptocurrencies (e.g., Bitcoin, Ethereum, Cardano, EOS, Monero) handle this problem by electing a \emph{proposer} who is responsible for producing a new block and sharing it with the network.
This proposer election happens via a distributed, randomized protocol chosen by the system designers. 

In Bitcoin, proposers are selected with probability proportional to the computational energy they have expended; this mechanism is called proof-of-work (PoW).
Under PoW, each node solves a computational puzzle of random duration; upon solving the puzzle, the node  relays its block over the underlying P2P network, along with proof that it solved the puzzle.
Due to the high energy cost of solving PoW puzzles (or \emph{mining}) \cite{energy}, a new paradigm recently emerged called \emph{proof-of-stake} (PoS).
Under PoS, a proposer is elected with  probability proportional to their stake in the system.
This election process happens at fixed time intervals.

When a node is elected proposer, its job is to propose a new block, which contains a hash of the previous block's contents.
Hence the proposer must choose where in the blockchain to append her new block. 
Most blockchains use a \emph{longest chain} fork choice rule, under which the proposer always appends her new block to the end of the longest chain of blocks in the proposer's local view of the blocktree. 
If there is no network latency and no adversarial behavior, this rule ensures that the blockchain will always be a perfect chain. 
However, in a network with random delays, it is possible that the proposer may not have received all blocks when she is elected. 
As such, she might propose a block that causes the blockchain to \emph{fork} (e.g.~Figure \ref{fig:3}). %\purple{(No figure here, more explanation may be needed)}.
In longest-chain blockchains, this forking is eventually resolved with probability 1 because one fork eventually overtakes the other.

Forking occurs in almost all major blockchains, and it implies that blockchains are often not chains at all, but \emph{blocktrees}.
 For many consensus protocols (particularly chain-based ones like Bitcoin's), forking reduces throughput, because blocks that are not on the main chain are discarded. 
It also has security implications; 
 even protocols that achieve good block throughput in the high-forking regime have thus far been prone to security vulnerabilities (which has been resolved in a recent work  \cite{prism}, which also guarantees low latency). 
Nonetheless, forking is a significant obstacle to {\em practical} performance in existing blockchains. 
%However, forking is difficult to mitigate, because it ultimately arises because of network latency. 
There are two common approaches to mitigate forking.
One is to improve the network itself, e.g. by upgrading hardware and routing.
% and/or routing algorithms.
This idea has been the basis for recent projects like the Falcon network \cite{falcon} and Bloxroute.
The other is to design consensus algorithms that tolerate network latency by making use of forked branches. 
Examples include GHOST \cite{ghost}, SPECTRE \cite{spectre}, and Inclusive/Conflux \cite{inclusive,conflux}.
In this paper, we  design a P2P protocol called \scheme~ that effectively reduces forking for a wide class of \emph{existing} consensus algorithms.

\bigskip 
\noindent {\bf Contributions.} We propose a novel probabilistic framework 
that allows one to formally investigate  the trade-off between 
the network delays and the throughput.  
We propose a new block proposal protocol 
to mitigate the forking due  to those network delays. 
We prove that when 
the proposer node 
polls $\ell$ randomly selected nodes for their local blocktree information, 
then it has the same effect as speeding up the communication network by a factor of $\ell$, 
thus reducing forking significantly. 
This is stated informally in the following and precisely in Theorem \ref{thm.dpolling}. 
\begin{theorem}[Informal] 
In a fully connected network with exponential network delays of mean $\delay$, 
let $L_\delay(t)$ denote the (random) number of blocks included in 
the longest chain at time $t$.
% for an arbitrary local attachment protocol 
%and an arbitrary block arrival process. 
For sufficiently small $\ell$,  under the proposed $\ell$-\scheme~polling, 
the resulting height of the longest chain is 
close to $L_{\delay/\ell}(t)$ 
 for any arbitrary block arrival process 
 and any  local attachment protocol. 
%G_{\delay}(t)$ denote the (random) local blocktree at a given node at time $t$, under a fixed, arbitrary local attachment protocol and an arbitrary block arrival process. The p.m.f. of $G_{\delay}(t)$ under $\ell$-\scheme~polling (for small $\ell$) is approximately equal to the p.m.f. of $G_{\delay/ \ell}(t)$ without $\ell$-\scheme~polling. 
\end{theorem}
These results hold without actually changing any network hardware, and they apply generally to any block arrival process or fork choice rule.  %Although the dynamics of polling are difficult to analyze in many cases, the proof for this statement is surprisingly simple, and relies only on reasoning about the time-evolution of a stochastic state matrix. 
In fact, we prove a significantly stronger statement; 
the \emph{entire blocktree probability mass function} changes to as if the network is faster by a factor of $\ell$, not just the downstream statistic of longest chain length. 
The analysis also has  connections to load balancing in balls-and-bins problems, %except the goal is to be as \emph{imbalanced} as possible; this 
which may be of independent interest. % to applied probability. 
We make the following three specific contributions:

\begin{enumerate} 
\item We propose a new probabilistic model for the evolution of a blockchain in 
proof-of-stake cryptocurrencies, 
where the main source of randomness comes from the network delay. 
This captures the network delays measured in real world P2P cryptocurrency networks   
% this model reflects randomness in 
% the underlying network, as measured in prior work 
\cite{DW13}.  
Simulations under this model explain the gap observed in real-world cryptocurrencies, 
between the achievable block throughput and 
the best block throughput possible in an infinite-capacity network. 
% how the block throughput depends on the network, and 
%We show empirically that there is a substantial gap between 
%block throughput under the proposed model and 
%the best block throughput possible in an infinite-capacity network. 
Our model differs from that of prior theoretical papers, which typically assume a worst-case network model that allows significant 
simplification in the analysis \cite{backbone,ghost}. 
We analyze the effect of average network delay on system throughput and  
provide  a lower bound on the block throughput. 

\item To mitigate forking due to network delays, 
we propose a new block proposal algorithm called $\ell$-\scheme, under which nodes poll $\ell$ randomly-selected nodes for their local blocktree information before proposing a new block. 
We show that for small values of $\ell$,  \scheme~ has approximately the same effect as if the {entire network} were a factor of $\ell$ faster. 

\item We  provide guidelines on how to implement \scheme~ in practice in order to provide robustness against several real-world factors, such as network model mismatch and adversarial behavior. 
% Our results show that \red{XYZ.}
\end{enumerate} 

\paragraph{Outline.} We begin by describing a stochastic model for blocktree evolution in Section \ref{sec:model}; we analyze the block throughput of this model in Section \ref{sec:3}. Next, we present \scheme~ and analyze its block throughput in Section \ref{sec:4}. Finally, we describe real-world implementation issues in Section \ref{sec:5}, such as how to implement polling and analyzing adversarial robustness.

%% file: sec2.tex
% !TEX root =  poll.tex
\section{Model}
\label{sec:model}

We propose a probabilistic model for blocktree evolution with two sources of randomness: 
randomness in the timing and the proposer of each new block, and 
the randomness in the delay in transmitting messages over the network.
The whole system is parametrized by 
the number of nodes $n$, 
average network propagation delay $\delay$, 
proposer waiting time $\tilde{\delay}$, 
and  number of concurrent proposers $k$.

% 
% ======================================================================================================== 
% 
\subsection{Modeling block generation} 
\label{sec:2a} 

We model block generation as a 
discrete-time arrival process, 
where the $t^{\text{th}}$ block is generated at  time $\gamma(t)$.
We previously discussed the election of a single proposer for each block; in practice, some systems elect multiple proposers at once to provide robustness if one proposer fails or is adversarial. 
Hence at time $\gamma(t)$,  $k$ nodes are chosen uniformly at random as {\em proposers},  
each of which proposes a distinct block. 
The index $t \in \mathbb Z^+$ is a positive integer, which we also refer to as {\em time} when it is clear from the context whether we are referring to $t$ or $\gamma(t)$. 
The randomness in choosing the proposers is independent across time and of other sources of randomness in the model. We denote the $k$ blocks proposed at time $t$ as $(t,1),(t,2),\ldots,(t,k)$. 
The block arrival process follows the distribution of a certain point process, which is independent of all other randomness in the model.

Two common block arrival process are Poisson and deterministic. 
Under a Poisson arrival process, $\gamma(t) - \gamma(t-1) \sim \mathsf{Exp}(\lambda)$ for some  constant $\lambda$, 
and $\gamma(t) - \gamma(t-1)$ is independent of $\{\gamma(i)\}_{i = 1}^{t-1}$. 
In proof-of-work (PoW) systems like Bitcoin, 
block arrivals are determined by independent attempts at solving a cryptographic puzzle, where each attempt has a fixed probability of success. 
With high probability, one proposer is elected each time a block arrival occurs (i.e., $k=1$), and   
 the arrival time can be modeled as a Poisson arrival process.

%This is more representative of proof-of-stake (PoS) algorithms. 
In many PoS protocols (e.g., Cardano, Qtum, and Particl), time is split into quantized intervals. 
Some protocols give each user a fixed probability of being chosen to propose the next block in each time interval, leading to a geometrically-distributed block arrival time.
If the probability of selecting \emph{any} proposer in each time slot is smaller than one, the expected inter-block arrival time will be greater than one, as in Qtum and Particl.
Other protocols explicitly designate one proposer per time slot (e.g., Cardano \cite{cardano_pos}).
Assuming all nodes are active, such protocols can be modeled with a deterministic interval process, $\gamma(t) = t$, for all $t \in \mathbb{N}$. 
The deterministic arrival process may even be a reasonable approximation for certain parameter regimes of protocols like Qtum and Particl. 
If the probability of electing any proposer in a time step is close to one, there will be at least one block proposer in each time slot with high probability, which can be approximated by a deterministic arrival process.
%Hence there would be a new block in each 
Regardless, our main results apply to arbitrary arrival processes $\gamma(t)$, including  geometric and deterministic. 

%    may only produce new blocks at geometric time intervals, if 
% if the 
%Some, like Cardano \cite{cardano}, already have a fixed inter-block time Qtum \cite{qtum}, and Particl \cite{particl},
%where block arrivals happen at regular time intervals, and multiple proposers can be elected at each time. 

%There is a subtlety to the application of $\ell$-polling which we delineate below. For clarity, consider the Bitcoin protocol where the  the proposer selection rule (conducted via ``mining", a form of proof of work)  is based on the block at the tip  of the longest chain -- in Bitcoin, mining is conducted on a hash of the block at the tip of the longest chain.  Thus in Bitcoin  if the result of  $\ell$-polling changes the longest chain, then the mining operation would no longer be valid (since the proof of work certificate is no longer valid on the latest block on the tip of the blockchain). On the other hand, several proof of stake cryptocurrencies avoid this issue by considering stake-based proposal mechanisms \cite{kiayias2017ouroboros,gilad2017algorand,pass2018thunderella} and $\ell$-polling applies directly. 
%\red{this is confusing, not clear what the implication is.}

%\red{why are we allowing poisson arrivals? I don't think a blockchain with Poisson arrivals would ever have $k>1$ }

% block generation -> proposer broadcast -> listener ... 
When a block $(t,i)$ is generated by a proposer, 
the proposer attaches the new block to one of the existing blocks, 
which we refer to as the \emph{parent block} of $(t,i)$. 
The proposer chooses this parent block according to a pre-determined rule called a {\em fork-choice rule}; we discuss this further in Section \ref{sec:2a}.  
Upon creating a block, the  proposer broadcasts a message containing the following information: 
\begin{align*}
M_{t,i} = (\text{Block } (t,i),\text{pointer to the parent block of } (t,i))
\end{align*} 
to all the other nodes in the system.   
The broadcasting process is governed by our network model, which is described in Section \ref{sec:2a}.
%Typical fork choice rules depend on the local view of the topology of all the existing blocks up to that point in time as explained in the following.  

{
In this work, we focus mainly on the PoS setting %(i.e., $\gamma(t)=t$) 
due to subtleties in the practical  implementation of \scheme~ (described in Section \ref{sec:4}). 
In particular, PoW blockchains require candidate proposers to choose a block's contents---including the parent block---\emph{before} generating the block.
But in PoW, block generation  itself takes an exponentially-distributed amount of time. 
Hence, if a proposer were to poll nodes before proposing, that polling information would already be (somewhat) stale by the time the block gets broadcast to the network.
In contrast, PoS cryptocurrencies allow block creation to happen \emph{after} a proposer is elected; hence polling results can be simultaneously incorporated into a block and broadcast to the network. 
Because of this difference, PoS cryptocurrencies  benefit more from \scheme~ than PoW ones.
% Although our theoretical results apply to arbitrary arrival processes $\gamma(t)$, we will focus mainly on the PoS setting in the paper. 
}

\bigskip
\noindent
{\bf Global view of the blocktree.}
Notice that the collection of all  messages forms a rooted tree, called the {\em blocktree}. 
Each node represents a block, and each directed edge represents a pointer to a parent block.
The root is called the \emph{genesis block}, and is visible to all nodes. 
All blocks generated at time $t=1$ point to the genesis block as a parent. 
The blocktree grows with each new block, since the block's parent must be an existing block in the blocktree;
since each block can specify only one parent, the data structure remains a tree. 
%No matter what fork choice rule is used, 
%a newly-generated block must point to a single, existing parent block, and hence this data structure is always a tree. 
Formally, we define the global blocktree as follows. 

\begin{definition}[Global tree]
We define the \emph{global tree} at time $t$, denoted as $G_t$, to be a graph whose edges are described by the set 
$
\{(\text{Block } (j,i),\text{pointer to the parent block of } (j,i)): 1\leq j\leq t, 1\leq i\leq k \}
$
with the vertices being the union of the genesis block and all the blocks indexed as $\{(j,i):1\leq j\leq t, 1\leq i\leq k\}$. 
\end{definition}

If there is no network delay in communicating the messages, 
then all nodes will have the same view of the blocktree.  
However, due to network delays and the distributed nature of the system, 
a proposer might add a block before receiving all the previous blocks. 
Hence, the choice of the parent node depends on the local view of the blocktree at the proposer node. 

\bigskip
\noindent
{\bf Local view of the blocktree.} 
Each node has its own local view of the blocktree, depending on which messages it has received.  
Upon receiving the message $M_{t,i}$, a node updates its local view as follows. 
If the local view contains the parent block referred in the message, then the block $t$ is attached to it. If the local view does not contain the parent block, then the message is stored in an \emph{orphan cache} until the parent block is received. 
Notice that $G_t$ is random and each node's local view is a subgraph of $G_t$.
% This communication takes a random amount of  time which is exponentially distributed with mean $\delay$, for each recipient node independently.

% For any \emph{local attachment protocol}, one can consider the union of the local trees at each node and define the global tree as follows. 

%{\bf Definition}:  The global tree $G_t$ at time $t$ consists of all messages of the form $(\text{Block } t,\text{pointer to Parent block of } t)$. 

% 
% ======================================================================================================== 
% 
\subsection{Network model and fork choice rule} 
\label{sec:2b} 

%We model the communication network as a complete graph, 
%where any node can communicate with any other node via a pairwise communication. 
We avoid modeling the topology of the underlying communication network by instead modeling the (stochastic) end-to-end delay of a message from any source to any destination node.
Stochastic network models have been studied for measuring the effects of selfish mining \cite{gobel2016bitcoin}  and blockchain throughput \cite{papadis2018stochastic}.  
We assume each block reaches a given node with delay distributed as an independent exponential random variable with mean $\delay$. 
This exponential delay captures the varying and dynamic network effects of real blockchain networks, as 
empirically measured in   \cite{DW13} on Bitcoin's P2P network. 
In particular, this exponential delay encompasses both network propagation delay and processing delays caused by nodes checking message validity prior to relaying it. 
These checks are often used to protect against denial-of-service attacks, for instance. 

When a proposer is elected to generate a new block at time $\gamma(t)$, 
%its local blocktree information is obtained instantaneously and 
she waits time $\tilde{\delay}\in [0,1)$ and decides on where to append the new block in its local blocktree. 
The choice of parent block is governed by the  fork choice rule. 
%How this attachment is done depends on the fork-choice rule. 
The most common one is the Nakamoto protocol (longest chain), though other fork choice rules do exist.
%{\bf Nakamoto protocol (longest chain).} 
When a node is elected as a proposer under the Nakamoto protocol (or longest chain rule), the node 
attaches the block to the leaf of the {\em longest chain} in the {\em local} blocktree. 
When there is a tie, %(i.e., if there are multiple longest chains), 
the proposer chooses one arbitrarily. 
Longest chain is widely-used, including in Bitcoin, ZCash, and Monero. 
The Nakamoto protocol belongs to 
 the family of \emph{local attachment protocols}, where 
 the proposer makes the decision on where to attach the block solely based on the snapshot of its \emph{local} tree at time $\gamma(t)+\tilde{\delay}$,  stripping away the information on the proposer of each block.
 In other words, we require that 
 the protocol be invariant to the identity of the proposers of the newly generated block. 
 We show in Section \ref{sec:4} that our analysis applies generally to all local attachment protocols. 
In practice, almost all blockchains use local attachment protocols.

 Notice that if $\delay$ is much smaller than the block inter-arrival time and all nodes obey protocol, 
 then the global blocktree $G_t$ is more likely to form a chain. 
 On the other hand, if $\delay$ is much larger than the block inter-arrival time, 
 then $G_t$ is more likely to be a star (i.e.~a depth-one rooted  tree). 
 To maximize blockchain throughput, it is desirable to design protocols that maximize the expected length of the longest chain of $G_t$. 
 Intuitively, a faster network infrastructure with 
 a smaller $\delay$ implies less forking. 
In this work, we are interested primarily in settings where $\delay$ is larger than the mean inter-block time. 
 This is admittedly not a conventional setting for existing blockchain systems, but a current trend in next-generation blockchains is to minimize block times and/or to run blockchains on increasingly unreliable networks (e.g., ad hoc networks, wireless networks, etc.). 
 In both settings, we may expect $\delay$ to be comparable to or larger than the block time. 
Hence our paper aims in part to understand the feasibility of operating blockchains in this regime.

\begin{comment}
 We can remove the storage assumption if we are taking large n scenarios, i.e. n >> (lkt)^2, in this case no proposer or polled node will  be a proposer or a polled node for a second time within a time interval t(with high probability), thus even if they store the polled information, it won't make any difference in our analysis. What do you think? - Ranvir 
\end{comment}

%% file: sec3.tex
% !TEX root =  poll.tex
\section{Block Throughput Analysis}
\label{sec:3}

A key performance metric in blockchains is \emph{transaction throughput}, or the number of transactions that can be processed per unit time. 
Transaction throughput is closely related to a property called \emph{block throughput}, also known as the main chain growth rate. 
Given a blocktree $G_t$, the length of the main chain $L(G_t)$ is defined as the number of hops from the genesis block to the farthest leaf. Precisely,
$$
L(G_t) \triangleq \max_{B \in \partial(G_t)} d(B_0, B),
$$
where $\partial(G_t)$ denotes the set of leaf blocks in $G_t$,  
and $d(B_0, B)$ denotes the hop distance between two vertices $B_0$ and $B$ in $G_t$.
We define \emph{block throughput} as $\lim_{t\to \infty} \Expect[L(G_t)]/t$.
% of this main chain is directly related to the \emph{throughput} of the system, 
% or the number of transactions that can be processed per unit time. 
Block throughput describes how quickly blocks are added to the blockchain; if each block is full and contains only valid transactions, then block throughput is  proportional to transaction throughput. 
In practice, this is not the case, since adversarial strategies like selfish mining \cite{selfish} can be used to reduce the number of valid transactions per block.
Regardless, block throughput is frequently used as a stepping stone for quantifying transaction throughput \cite{ghost,backbone,prism}.

For this reason, a key objective of our work is to quantify block throughput, both with and without polling. 
%We begin by quantifying the main-chain growth rate without polling. 
We begin by studying block throughput  without polling 
under the Nakamoto protocol  fork-choice rule, as in Bitcoin. 
This has been previously studied in \cite{ghost,backbone,prism},  
under a simple network model where there is a fixed deterministic delay between any pair of nodes. 
This simple network model is justified by arguing that 
if all transmission of messages are guaranteed to arrive within a fixed maximum delay $d$, 
then the worst case of block throughput  happens when all transmission have delay of exactly $d$. 
%Although other papers have studied bounds on Bitcoin's block 
%throughput \cite{ghost,backbone,prism}, their results make worst-case network assumptions that facilitate analysis.
%In particular, they assume that all transactions arrive within a (typically known) time threshold. 
%However, in many network applications, we are more interested in the average throughput of a system  than the worst-case throughput. 
Such practice ignores all the network effects, for the sake of tractable analysis. 
In this section, we focus on capturing such network effect on the block throughput. 
We ask the fundamental question of 
how block throughput depends on the average network delay, under a more realistic network model where 
each communication is a realization of a random exponential variable with average delay $\delay$. 
%Another important difference is that in Bitcoin, blocks arrive according to a Poisson process, 
%whereas in PoS systems, blocks come at (roughly) fixed time intervals.
In the following (Theorem \ref{thm:len}), we provide a  lower bound on the block throughput, under the more nuanced network model from Section \ref{sec:model}, 
and Nakamoto protocol fork-choice rule.
This result holds for a deterministic arrival process.
We refer to a longer version of this paper  \cite{longer} for a proof. 
%A proof is provided in Section \ref{sec:len_proof}. 
%The following theorem makes this precise. 
%{\bf a paragraph on how security depends on length.}
%
%{\bf past work has focused on worst case network model.} 

%define throughput of a 
%In this section, we lower bound the growth rate of the main chain under our network model.

\begin{theorem}
	\label{thm:len}
	Suppose there is a single proposer ($k=1$) at each discrete time, $\gamma(t)=t \in\{1,2,\ldots\}$, with no waiting time ($\tilde{\delay}=0$).  
	For any number of nodes $n$, any time $t$,  any average delay $\delay$, and $C_\delay =e^{\frac{-1}{\delay}}$, under the Nakamoto protocol, we have that 
	\begin{align*}
		\frac{\Expect[L_{\mathsf{Chain}}(G_t)]}{t} \;\; \geq \;\; \exp\pth{\frac{-C_\delay}{(1-C_\delay)^2}} \;. 
	\end{align*}
%	(Proof in Appendix \ref{})
%	\red{Does this result hold for all $t$? Yes it does}
\end{theorem}

\begin{figure}[t]
\centering
	\includegraphics[width = 0.6\linewidth]{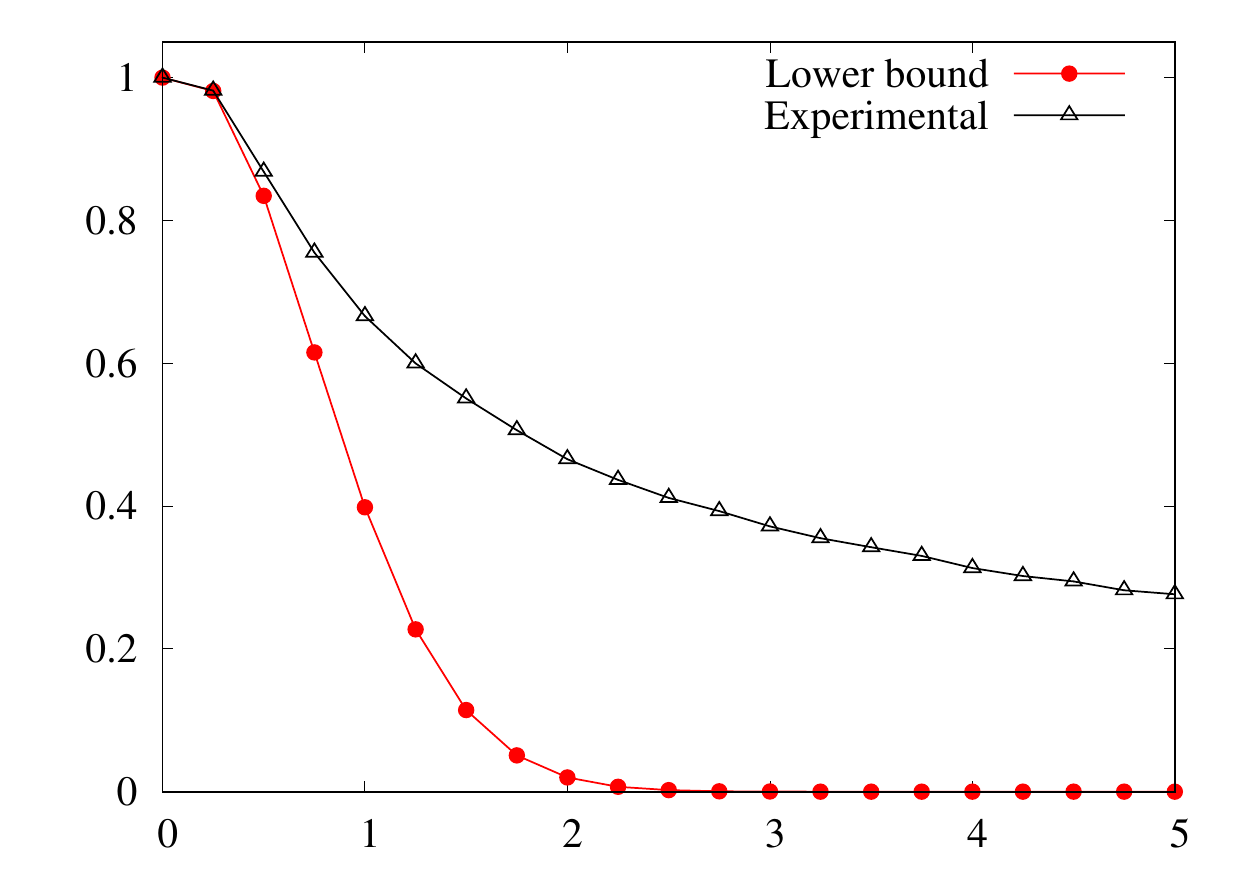}
	\put(-110,-5){average network delay $\delay$}
	\put(-150,5){\rotatebox{90}{average block throughput}}
	\caption{Block throughput vs. average network delay for an inter-block time of 1 time unit. }
	\label{fig:len}
\end{figure}

Notice that trivially, ${\Expect[L_{\mathsf{Chain}}(G_t)]}/{t} \leq 1$, with equality when there is no network delay, $\delay=0$. 
%For example, when $\delay=1$, Theorem \ref{thm:len} implies that ${\Expect[L_{\mathsf{Chain}}(G_t)]}/{t} \geq 0.39$. 
Theorem~\ref{thm.chainhighprob} and our experiments in Figure \ref{fig:len} suggest that Theorem \ref{thm:len} is tight when $\delay \ll 1$. Hence there is an (often substantial) gap between the realized block  throughput and the desired upper bound.
%The gap between realized block throughput and the optimal throughput of 1 occurs because of network delays, and is the focus of this work. 
This gap is caused by network delays;
since proposers may not have an up-to-date view of the blocktree due to network latency, they may append to blocks that are not necessarily at the end of the global main chain,
thereby causing the blockchain to \emph{fork}.

One goal is to obtain a blocktree with no forking at all, i.e., a perfect blockchain  with $L_{\mathsf{Chain}}(G_t) = t$.
Setting $\exp\pth{{-C_\delay}/{(1-C_\delay)^2}} = 1- {1}/{t}$, which implies that $\Expect[L_{\mathsf{Chain}}(G_t)] \geq t-1$, we obtain that $\delay = \Theta(\frac{1}{\log t})$. The following result shows that if $\delay = O(\frac{1}{\log t})$, then $L_{\mathsf{Chain}}(G_t) = t$ with high probability. 
%In other words, the blocktree is a chain. 

%\setlength{\abovedisplayskip}{3pt}
%\setlength{\belowdisplayskip}{3pt}

\begin{theorem}\label{thm.chainhighprob}
Fix a confidence parameter $\delta \in (0,1),k =1, \gamma(t) = t$. For the Nakamoto protocol, if 
\begin{align}
\frac{1}{\delay} \geq  \frac{ \left( \ln t - \ln \ln \frac{1}{\delta} \right)}{1-\tilde{\delay}},
\end{align} 
then the chain $\text{Gen}-1-2-\ldots-t$ happens with probability at least $\delta - o(1)$ as $t\to \infty$ and $n \gg t^2$. 

Conversely, when $n\gg \left( \delay t \ln t \right)^2$ and 
\begin{align}
\frac{1}{\delay} \leq \frac{\left( \ln t - \ln \ln \frac{1}{\delta} \right)}{1-\tilde{\delay}},
\label{eq:fork_whp}
\end{align}
then the chain $\text{Gen}-1-2-\ldots-t$ happens with probability at most $\delta + o(1)$ as $t\to \infty$. Here $\gg$ ignores the dependence on the parameter $\delta$, which is fixed throughout. 
\end{theorem}
The proof is included in Section  \ref{sec:highprobaproof}. 
This result shows the prevalence of forking.  
For example, if we conservatively use Bitcoin's parameters settings, taking  $\delay=0.017$, $\tilde \delay=0$, and $\delta=0.01$, equation \eqref{eq:fork_whp} implies that for $t \gtrsim 5$ blocks, forking occurs with high probability.
Hence forking is pervasive even in systems with parameters chosen specifically to avoid it.

A natural question is how to reduce forking, and thereby increase block throughput.
To this end, we next  introduce a blockchain evolution protocol called \scheme, that effectively reduces forking without changing the system parameter $\delay$, which is determined by network bandwidth. 
%The $\ell$-polling strategy works as follows: upon arrival of a block $(t,i)$, the proposer of block $(t,i)$ selects $\ell-1$ nodes in the network uniformly at random, and inquires about their local tree. The proposer aggregates the information from the $\ell -1$ other nodes and makes a decision on where to attach block $(t,i)$ based on the \emph{local attachment protocol} it follows. One key observation is that there is no conflict between the local trees of each node, so the $\ell$-polling strategy simply merges totally $\ell$ local trees into a union tree. We assume that when $\ell$-polling happens, the polling requests arrive at the polled nodes instantaneously, and it takes the proposer node time $\tilde{\delay}$ to make the decision on where to attach the block. We also assume that each node processes the additional polled information real time with no storage to simplify the analysis. In other words, the information a node polled at time $t$ is forgotten at time $t'>t$.

%% file: sec4.tex
% !TEX root =  poll.tex
\section{$\ell$-\scheme} 
\label{sec:4}

To reduce forking and increase block throughput, we propose $\ell$-\scheme, which works as follows: upon arrival of a block $(t,i)$, the proposer of block $(t,i)$ selects $\ell-1$ nodes in the network uniformly at random, and inquires about their local tree.\footnote{We use the name \scheme~ to refer to the general principle, and $\ell$-\scheme~to refer to an instantiation with polling parameter $\ell$.} The proposer aggregates the information from the $\ell -1$ other nodes and makes a decision on where to attach block $(t,i)$ based on the \emph{local attachment protocol} it follows. 
One key observation is that there is no conflict between the local trees of each node, 
so the \scheme~ strategy simply merges totally $\ell$ local trees into a single  tree with union of all the edges in the local trees that are polled. 
Note that we poll $\ell-1$ nodes, 
such that a total $\ell$  local trees are contributing, 
as the proposers own local tree also contributes to the union. 

We assume that when \scheme~polling happens, the polling requests arrive at the polled nodes instantaneously, and it takes the proposer node time $\tilde{\delay}$ to make the decision on where to attach the block. 
The instantaneous polling assumption is relaxed in Section \ref{sec:5}. 
Recall that in our model, $\delay$ accounts for both network delay and processing delays. 
In live blockchain P2P networks, a substantial fraction of block propagation delays originate from the processing (e.g. validity checks) done by each node before relaying the block.
These delays could grow more pronounced for blockchains with more complex processing requirements, such as smart contracts.
Since these computational checks are not included in the polling process, the polling delay can be much smaller than the overall network propagation delay. 
To simplify the analysis, we also assume that each node processes the additional polled information in real time, but does not store the polled information. %with no storage, 
 In other words, the information a node obtains from polling at time $t$ is forgotten at time $t'>t + \tilde{\delay}$.
This modeling choice is made to simplify the analysis; it results in a lower bound on the improvements due to polling since nodes are discarding information.
% \red{Wouldn't it be forgotten after time $t'>t+\tilde \delay$ if the node needs $\tilde \delay$ to decide? Yes, here we can add $t'>t+\tilde{\delay}$}
In practice, network delay affects polling communication as well, and we investigate experimentally 
these effects in Section \ref{sec:delay}. 

To investigate the effect of polling on the blockchain, 
we define appropriate events on the probabilistic model of block arrival and block tree growth. 
We denote $X\sim \mathsf{Exp}(\lambda)$ an exponential random variable with probability density function $p_X(t) = \lambda e^{-\lambda t} \mathds{1}(t\geq 0)$, and define set $[m] \triangleq \{1,2,\ldots,m\}$ for any integer $m\geq 1$. 
For a message
\begin{align*}
M_{j,i} \;\; =\;\; (\text{Block }(j,i), \text{point to the parent block of }(j,i)), 
\end{align*}
denote its arrival time to node $m$ as $R_{(j,i),m}$. If $m$ is the proposer of block $(j,i)$, then $R_{(j,i),m} = \gamma(j)+\tilde{\delay}$. If $m$ is not the proposer of block $(j,i)$, then $R_{(j,i),m} = \gamma(j) +\tilde{\delay}+ B_{(j,i),m}$, where $B_{(j,i),m} \sim \mathsf{Exp}(1/\delay)$. It follows from our assumptions that the random variables $B_{(j,i),m}$ are mutually independent for all $1\leq j\leq t, 1\leq i\leq k, 1\leq m\leq n$. We also denote the proposer of block $(j,i)$ as $m_{(j,i)}$. 
To denote polled nodes, we also write $m_{(j,i)}$ as $m_{(j,i)}^{(1)}$, and denote the other $\ell-1$ nodes polled by node $m_{(j,i)}$ as $m_{(j,i)}^{(2)}, m_{(j,i)}^{(3)}, \ldots, m_{(j,i)}^{(\ell)}$. 

When block $(j,i)$ is being proposed, we define the following random variables. Let random variable
\begin{align}
e_{j,i,l,r} = \begin{cases} 1 & \text{if by the time }(j,i)\text{ was proposed, node } \\
& \quad  m_{(j,i)}^{(l)}\text{ already received block }r \\ 0 & \text{otherwise}   \end{cases}
\end{align}
Here $j\in [t], i\in [k], l\in [\ell], r \in \{(a,b):a\in [j-1], b\in [k]\}$. For any $r = (a,b)$, we denote $r[1] = a, r[2] = b$.  

Since we will aggregate the information from the total $\ell$ nodes whenever a proposer proposes, we also define
$e_{j,i,r} =1-\prod_{l=1}^\ell (1-e_{j,i,l,r})$ as the 
event that when $(j,i)$ was proposed, at least one node  $m_{(j,i)}^{(l)}$ 
 has received block $r$.  
The crucial observation is that when the proposer tries to propose block $(j,i)$, the complete information it utilizes for decision is the collection of random variables
\begin{align}
\{e_{j,i,r}: r[1]\in [j-1], r[2]\in [k]\}. 
\end{align}

The global tree at time $\gamma(t)+\tilde{\delay}$, denoted as $G_t$, is a tree consisting of $k t + 1$ blocks including the Genesis block. We are interested in the distribution of the global tree $G_t$. To illustrate how to compute the probability of a certain tree structure, we demonstrate the computation through an example where $k = 1, t = 3$, and $\ell = 1$. %, \tilde{\delay} = 0$. 

\begin{figure}[h]  
\centering 
\includegraphics[width=1.7in]{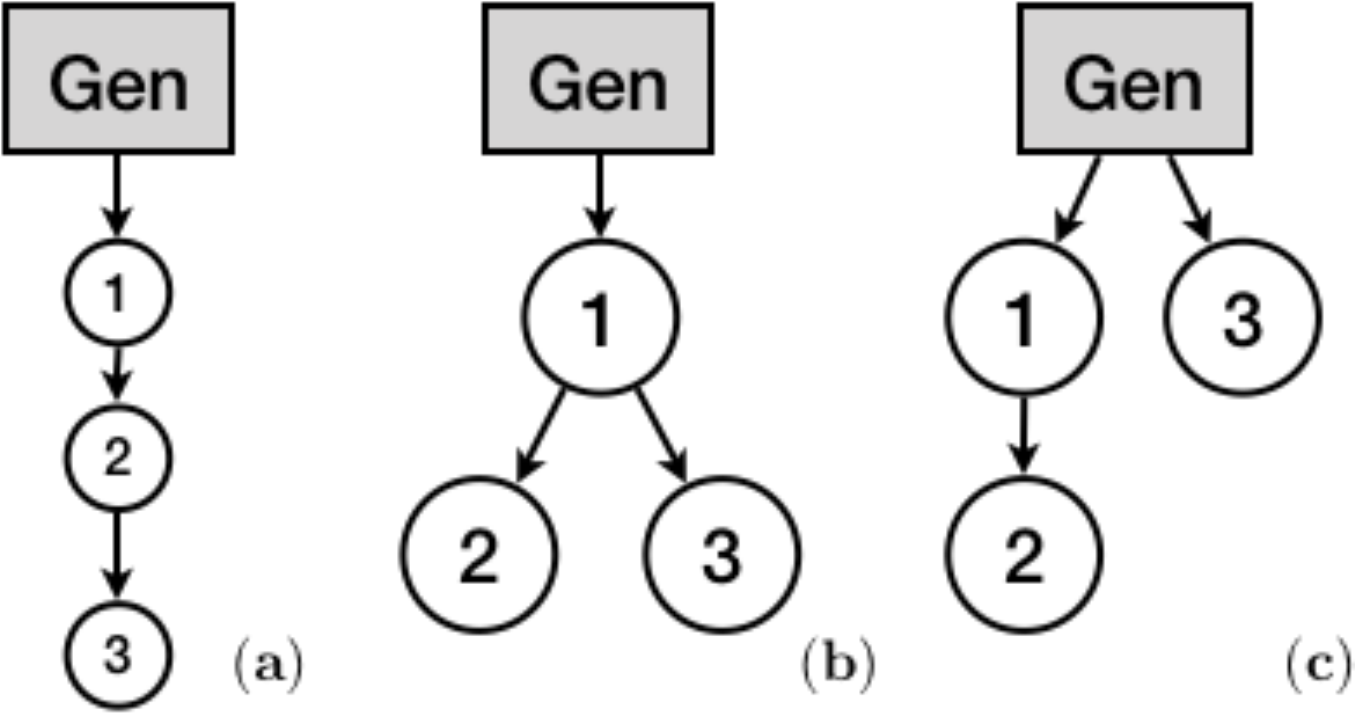}
\caption{Examples of $G_3$ with varying structures.}
\label{fig:3}
\end{figure}

For simplicity, we denote $e_{j,i,(r[1],r[2])}$ as $e_{j,r[1]}$ since for this example $k = 1$. 
%\red{Is this backwards? Isn't $r[1]=1$ always since $k=1$? $l$ is the index of polled nodes, which here is always 1. Note that the definition of $e_{j,i,r}$ already ignores $l$. $r$ in fact has two components. In fact here $r[2]$ is always 1, but $r[1]$ is the index of previous blocks, and could be 1 or 2 here since we in total have proposed 3 blocks. }
The probability of some of the configurations of $G_3$ in Figure~\ref{fig:3}a can be written as
\begin{align*}
\prob{G_3= \text{Figure \ref{fig:3}a}} & = \mathbb{P}\left( e_{2,1} = 1, e_{3,1} = 1, e_{3,2} = 1 \right )\;, \\
\prob{G_3= \text{Figure \ref{fig:3}b}} & = \mathbb{P}\left( e_{2,1} = 1, e_{3,1} = 1, e_{3,2} = 0 \right )\;, \text{ and } \\
\prob{G_3= \text{Figure \ref{fig:3}c}} & = \mathbb{P}\left( e_{2,1} = 1, e_{3,1} = 0 \right ) \;.
\end{align*}
Note that for the event in Figure \ref{fig:3}, it does not matter whether node 
$m_{(3,1)}$ has received block $(2,1)$ or not, as the parent of that block is missing in $m_{(3,1)}$'s local tree. 
Block $(2,1)$ is therefore not included in the local tree of node $m_{(3,1)}$ at that point in time. 

\subsection{ Main result}
\label{sec:main}
Under any local attachment protocol $\mathcal{C}$ and any block arrival distribution, 
the event that $E_{C,t,g} = \{G_t = g\}$ depends on 
the random choices of proposers and polled nodes, 
$\{m_{(j,i)}^{(l)}:j\in [t],i\in [k],l\in [\ell]\}$, and 
the messages received at those respective nodes, 
$\{e_{j,i,r}: j\in [t],i \in [k], r[1]\in [j-1], r[2]\in [k]\}$, and 
some additional outside randomness on the network delay and the block arrival time.  
The following theorem characterizes the distribution of $G_t$ on the system parameters $t, \delay, \ell,\tilde{\delay}$ for a general local attachment protocol $\mathcal{C}$ (including the longest chain protocol). 
We provide a proof in Section \ref{sec:polling_proof}.

\begin{theorem}
\label{thm.dpolling}
For any local attachment protocol $\mathcal{C}$ and any inter-block arrival distribution,
define random variable $\tilde{G}_t$ which takes values in the set of all possible structures of tree $G_t$ such that \footnote{The random variable $\tilde{G}_t$ is well defined, 
since the protocol $\mathcal{C}$ is assumed not to depend the identity of the proposer of each block. Hence, the conditional expectation is identical conditioned on each specific $\{m_{(j,i)}^{(l)}: j\in [t], i\in [k], l\in [\ell]\}$ whenever all $tk\ell$ nodes in it are distinct. }
\begin{align} 
& \mathbb{P}(\tilde{G}_t = g) \;\; \triangleq \nonumber \\ 
& \;\;   \mathbb{E}\Big[\mathds{1}(E_{C, t,g}) \big| \big\{m_{(j,i)}^{(l)}\big\}_{j\in [t],i\in [k],l\in [\ell]}\text{ are distinct} \Big]. \label{eqn.tildegdefinition}
\end{align}

We have the following results: 
\begin{enumerate}
\item[$(a)$] There exists a function $F$ independent of all the parameters in the model such that for any possible tree structure $g$,  
\begin{align}
	\mathbb{P}(\tilde{G}_t = g) \;\; =\;\; F\Big(\, \frac{\delay}{\ell}, \tilde{\delay}, g, \mathcal{C}\,\Big)\;.
	\label{eq:ell}
\end{align}
\item[$(b)$] The total variation distance between the distribution of $G_t$ and $\tilde{G}_t$ is upper bounded:
\begin{align}
	\mathsf{TV}\big(\,P_{G_t}, P_{\tilde{G}_t}\,\big) \;\; \leq\;\; \frac{(\ell k t )^2}{2n} \;. 
	\label{eq:tv}
\end{align}
\end{enumerate} 
\end{theorem}

\begin{figure}[!htb]
	\centering
	\includegraphics[width = 0.6\linewidth]{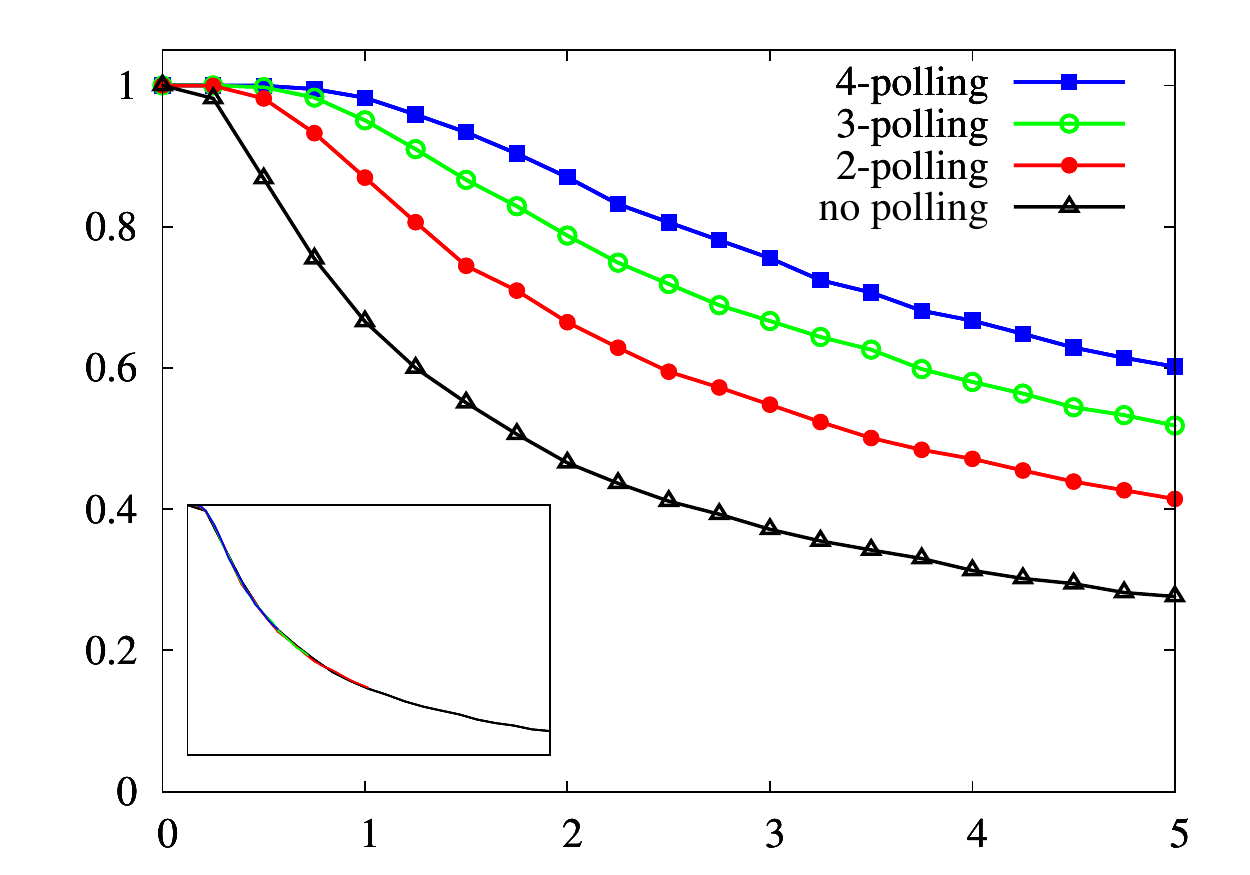}
	\put(-115,-5){mean network delay $\delay$}
	\put(-145,5){\rotatebox{90}{Average block throughput}}
	\caption{Comparing the average block throughput for various choices of $\ell$  
	confirms the theoretical prediction that $\ell$-\scheme~ effectively speeds up the network by a factor of $\ell$; 
	all curves are indistinguishable when x-axis is scaled as $\delay/\ell$ as shown in the inset.
	}
	\label{fig:main}
\end{figure}

In the definition in Eq.~\eqref{eqn.tildegdefinition}, 
we condition on the event that all proposers and polled nodes are distinct. 
This conditioning ensures that all received blocks $e_{j,i,l,r}$'s at those nodes are independent over time $j$. 
This in turn allows us to capture the precise effect of $\ell$ in the main result in Eq.~\eqref{eq:ell}.
Further, the bound in Eq.~\eqref{eq:tv} implies that such conditioning is not too far from 
the actual evolution of the blockchains, as long as the number of nodes are large enough: $n\gg (\ell k t)^2$.
While this condition may seem restrictive since $t \to \infty$, notice that in practice, many blockchains operate in \emph{epochs} of finite duration, such that the state of the blockchain is finalized between epochs \cite{casper,kiayias2017ouroboros}. 
Finalization means that the system chooses a single fork, and builds on the last block of that fork in the subsequent epoch. 
Hence, the above condition can be physically met with finite $n$. 
Moreover, in practice, $n$ need not be so large, 
as we show in Figure \ref{fig:main}. 
Even with $n=10,000 <  (\ell k t)^2 = 160,000$ for 4-polling, 
the experiments support the predictions of Theorem \ref{thm.dpolling}. 

The main message of the above theorem is that $\ell$-\scheme~effectively reduces the network delay by a factor of $\ell$.  
For {\em any} local attachment protocol and \emph{any} block arrival process, 
up to a total variation distance of $(\ell k t)^2/n$, 
the distribution of the evolution of the blocktree 
with $\ell$-\scheme~ is the same as the distribution 
of the evolution of the blocktree with no polling, but with a network that is $\ell$ times faster. 
We confirm this in numerical experiments (plotted in Figure \ref{fig:main}), 
%Theorem~\ref{thm.dpolling} implies that as long as $n \gg (\ell k t)^2$, the effect of $\ell$-polling can be approximately described as changing the delay parameter $\delay$ to $\delay/\ell$. 
for a choice of $\tilde{\delay}=0$, 
$n=10,000$, $k=1$, $t=100$, $\gamma(t)=t$, and the longest chain fork choice rule. 
In the inset we show the same results, but scaled the x-axis as $\delay/\ell$. 
As predicted by Theorem \ref{thm.dpolling}, 
the curves converge to a single curve, and are indistinguishable from one another.  
We used the network model from Section \ref{sec:2b}.
%We observe from the plot that Main chain length for no polling at delay D almost same as the Main chain length at delay D/l if we do l-polling. For example, $L(G_{100})$ for no polling at $D=1$ is 66.58, which is almost same as the $L(G_{100})$ for $2$ polling at $D=2$ (66.47), $L(G_{100})$ for $3$ polling at $D=3$ (66.65),$L(G_{100})$for $4$-polling at $D=4$ (66.69). This experimentally verifies theorem X.
%Moreover, the performance increase is quite significant as the $L(G_100)$ for $D=1$ increases significantly from 66.58(no-polling) to 86.96(2-polling), 95.06(3-polling), 98.27(4-polling).

Without polling, the throughput degrades quickly as the network delay increases. 
This becomes critical as we try to scale up PoS systems; blocks should be generated more frequently, pushing network infrastructure to its limits. 
With polling, we can achieve an effective speedup of the network without investing resources on hardware upgrades.
Note that in this figure, we are comparing the average block throughput, which is the main property of interest. 
We make this connection between the throughput and $\ell$ precise in the following. 
Define $L_{\mathsf{Chain}}(G_t)$ to be the length of the longest chain in $G_t$ excluding the Genesis block. 
Throughput is defined as ${\mathbb E}[L_{\mathsf{Chain}}(G_t)]/t$. 
We have the following Corollary of Theorem~\ref{thm.dpolling}. 

\begin{corollary}\label{cor.thm1}
There exists a function $L(  {\delay}/{\ell}, \tilde{\delay},\mathcal{C})$ independent of all the parameters in the model such that 
\begin{align}
\Big| \, \mathbb{E}[L_{\mathsf{Chain}}(G_t)] - \mathbb{E}\Big[ L\Big(  \frac{\delay}{\ell}, \tilde{\delay},\mathcal{C}\Big)\, \Big]\,  \Big|  \; \leq\;  
	\frac{t (\ell k t)^2 }{2n}. 
\end{align}
%Here the expectation in $\mathbb{E}[ L(\{\gamma(i)\}_{i =1}^t, \frac{D}{\ell}, \tilde{D},\mathcal{C})]$ is over the randomness of the arrival process $\{\gamma(i)\}_{i =1}^t\}$. 
\end{corollary}

In other words, in the regime that $n \gg t^3 (k\ell)^2$, the expectation of the length 
of the longest chain depends on the delay parameter $\delay$ and the polling parameter 
$\ell$ only through their ratio ${\delay}/{\ell}$. 
Hence, the block throughput enjoys the same polling gain, as the distribution of the resulting block trees.

%
% ---------------------------------------------------------------------------------------------------------------------------------------------------------------------
%
%\subsection{}
%\label{sec:extreme}
%

%
% ---------------------------------------------------------------------------------------------------------------------------------------------------------------------
%
\subsection{Connections to balls-in-bins example}
\label{sec:balls}

In this section, we give a brief explanation of the  balls-in-bins problem and the power of two choices in load balancing. 
We then make a concrete connection between the blockchain problem and the power of $\ell$-polling in information balancing.

In the classical balls-in-bins example, 
we have $t$ balls and $t$ bins, and 
we sequentially throw each ball into a uniformly randomly selected bin. 
Then, the maximum loaded bin has load (i.e.~number of balls in that bin) 
scaling as $\Theta\left( {\log t}/{\log \log t} \right)$~\cite{mitzenmacher2005probability}. 
The result of \emph{power of two choices} states that if every time we select $\ell$ ($\ell \geq 2$) 
bins uniformly at random and throw the ball into the \emph{least} loaded bin, 
the maximum load enjoys an near-\emph{exponential} reduction to 
$\Theta\left( {\log \log t}/{\log \ell} \right)$~\cite{mitzenmacher2005probability}. 

Our polling idea is inspired by this power of two choices in load balancing. 
We make this connection gradually more concrete in the following. 
First, consider the case when 
the underlying network is extremely slow such that no broadcast of the blocks is received. 
When there is no polling, 
each node is only aware of its local blockchain consisting of only those blocks it generated. 
There is a one-to-one correspondence to the balls-in-bins setting, as 
blocks (balls) arriving at each node (bin) build up a load (local blockchain). 
When there are $t$ nodes and $t$ blocks, 
then it trivially follows that the length of the longest chain scales as $\Theta( {\log t}/{\log \log t})$, 
when there is no polling. 

The main departure is that in blockchains, the goal is to maximize the 
length of the longest chain (maximum load).   
This leads to the following fundamental question in the balls-in-bins problem, which has not been solved, to the best of our knowledge.
%has not been addressed in the literature to the best of our knowledge. 
If we throw the ball into the {\em most} loaded bin among $\ell$ randomly chosen bins at each step, 
how does the maximum load scale with $t$ and $\ell$? 
That is, if one wanted to maximize the maximum load, leading to load unbalancing, 
how much gain does the power of $\ell$ choices give? 
We give a precise answer in the following.

\begin{theorem}\label{thm.ballsandbins}
Given $t$ empty bins and $t$ balls, we sequentially allocate balls to bins as follows. For each ball, we select uniformly at random $\ell$ bins, and put the ball into the maximally-loaded bin among the $\ell$ chosen ones. Then, the maximum load of the $t$ bins after the placement of all $t$ balls is at most
\begin{align}
C\cdot \ell \cdot \frac{\log t}{\log \log t}
\end{align}
with probability at least $1-\frac{1}{t}$, where $C>0$ is a universal constant. 
\end{theorem}

We refer to a longer version of this paper  \cite{longer} for a proof. %We provide a proof in Section \ref{sec:balls_proof}. 
%This result is analogous to Theorem 5 from \cite{redlich2013unbalanced}, which uses a coupling argument to show the result; we instead directly bound the load of the bin.  
This shows that 
 the gain of $\ell$-polling in maximizing the maximum load is linear in $\ell$. 
Even though this  is not as dramatic as  the exponential gain of the load balancing case, 
this gives a precise characterization of the gain in the throughput of $\ell$-\scheme~ in blockchains when $\delay\gg1$.  
This is under a slightly modified protocol where 
the polling happens in a bidirectional manner, such that 
the local tree and the newly appended block of the proposer are also sent to the polled nodes.

For moderate to small $\delay$ regime, which is the operating regime of real systems, 
blocktree evolution is connected to a generalization of the balls-in-bins model. 
Now, it is as if the balls are copied and broadcasted to all other bins over a communication network. 
This is where the intuitive connection to balls-and-bins stops, as we are storing the information in a specific data structure that we call blocktrees. 
However, we borrow the terminology from `load balancing', and refer to the effect of polling as `information balancing', 
even though load balancing refers to {\em minimizing} the maximum load, whereas 
information  balancing refers to {\em maximizing} the maximum load (longest chain) by balancing the information throughout the  nodes using polling.

%% file: sec5.tex
% !TEX root =  poll.tex
\section{System and implementation issues}
\label{sec:5}

We empirically verify the robustness of our proposed protocol under various issues 
that might come up in a practical implementation of $\ell$-\scheme. 
Our experiment consists of $n$ nodes connected via a network which emulates the end to end delay as an exponential distribution; this model is inspired by the measurements of the Bitcoin P2P network made in   \cite{DW13}.  

Each of the $n$ nodes maintains a local blocktree which is a subset of the global blocktree. 
We use  a deterministic block arrival process with $\gamma(t)=t$, i.e. 
we assume a unit block arrival time which is also termed as an epoch in this section. 
This represents an upper bound on block arrivals in real-world PoS systems, where blocks can only arrive at fixed time intervals.
%We keep concurrent number of 
At the start of arrival $t$, $k$ proposers are chosen at random and 
each of these proposers proposes a block. 

When there is no polling, each proposer chooses the most eligible block from its blocktree to be a parent to the block it is proposing, based on the fork choice rule. 
In the case of $\ell$-\scheme, the proposer sends a pull message to $\ell-1$ randomly chosen nodes, and these nodes send their block tree back to the proposer. 
The proposer receives the block trees from the polled nodes after a delay $\tilde{\delay}$, and 
updates her local blocktree by taking the union of all received blocktrees. 
The same fork choice rule is applied to  decide the parent to the newly generated block. 
In all  experiments, Nakamoto longest chain fork choice rule is used. Experiments are run for $T=100$ time epochs on a network with $n=10,000$ nodes with $k=1$. 

%There is a subtlety to the application of $\ell$-polling which we delineate below. For clarity, consider the Bitcoin protocol where the  the proposer selection rule (conducted via ``mining", a form of proof of work)  is based on the block at the tip  of the longest chain -- in Bitcoin, mining is conducted on a hash of the block at the tip of the longest chain.  Thus in Bitcoin  if the result of  $\ell$-polling changes the longest chain, then the mining operation would no longer be valid (since the proof of work certificate is no longer valid on the latest block on the tip of the blockchain). On the other hand, several proof of stake cryptocurrencies avoid this issue by considering stake-based proposal mechanisms \cite{kiayias2017ouroboros,gilad2017algorand,pass2018thunderella} and $\ell$-polling applies directly. 
%\red{this is confusing, not clear what the implication is.}

%This shows that the chain growth rate is constant, and the system stabilizes fast.
%\red{Is this last sentence saying anything interesting? Also, what is the point of Figure 4?}
%\purple{This statement shows transient behavior and might be useful if we decide to provide some theorem for the upper bound(not finalized yet), as it may use constant growth rate assumption. Its not that important and we can remove it if we want to.}

\subsection{Effect of polling delay}
\label{sec:delay}

In reality, there is delay between initializing a poll request and receiving the blocktree information. 
We expect polling delay to be smaller than the delay of the P2P relay network because polling communication is point-to-point rather than occurring through the P2P relay network.
% and
%(2) we find that empirically, polled nodes need not send the entire blocktree to obtain the benefits of $\ell$-polling.  
%one can expect it to be faster than the delay of the standard block broadcasts. 
%We assume that this communication and further processing takes a random time which follows an $\mathsf{Exp}(\frac{1}{0.1D})$ distribution.
%To motivate point (2), notice that our model of $\ell$-polling % studied in this paper  
%implicitly assumes the complete local blocktrees are synched. 
%This may unnecessarily waste network resources. 
%For efficient bandwidth usage, we consider $(\ell,b)$-polling, where 
%the polled nodes only send the blocks that were generated between times $t-1$ and $t-b$. 
%% This limits the syncing time as the maximum information a polled node will have to send is $b$ blocks. 
%In Figure \ref{fig:latest5}, 
%we compare the performance of 2-polling and (2,5)-polling. 
%The experiments suggest that  
%comparable performance can be achieved with a choice of $b$ that is small, 
%and that choice increases with  network delay. 
%Hence, we model the polling delay as smaller than the P2P relay network's mean delay.
%%Thus, this is a good algorithm to bound communication required for polling. $b$ needs to be adjusted according to the network delay, higher network delay requires higher $b$.
To	 understand the effects of polling delay, 
we ran simulations in which a proposer polls $\ell-1$ nodes at the time of proposal, 
and  each piece of polled information arrives after time $\tilde{\delay}_1$, $\tilde{\delay}_2$, .., $\tilde{\delay}_{\ell-1} \sim \mathsf{Exp}(\frac{1}{0.1\,\delay})$. 
The proposer determines the pointer of the new block when all polled messages are received. 
%Thus we model $$\tilde{\delay}= \max_{i \in \{1,2,..,\ell-1\}} \tilde{\delay_i}$$
%The proposer proposes at time $t+\tilde{\delay}$. 

\begin{figure*}[!htb]
\minipage{0.32\textwidth}
  \includegraphics[width=\linewidth]{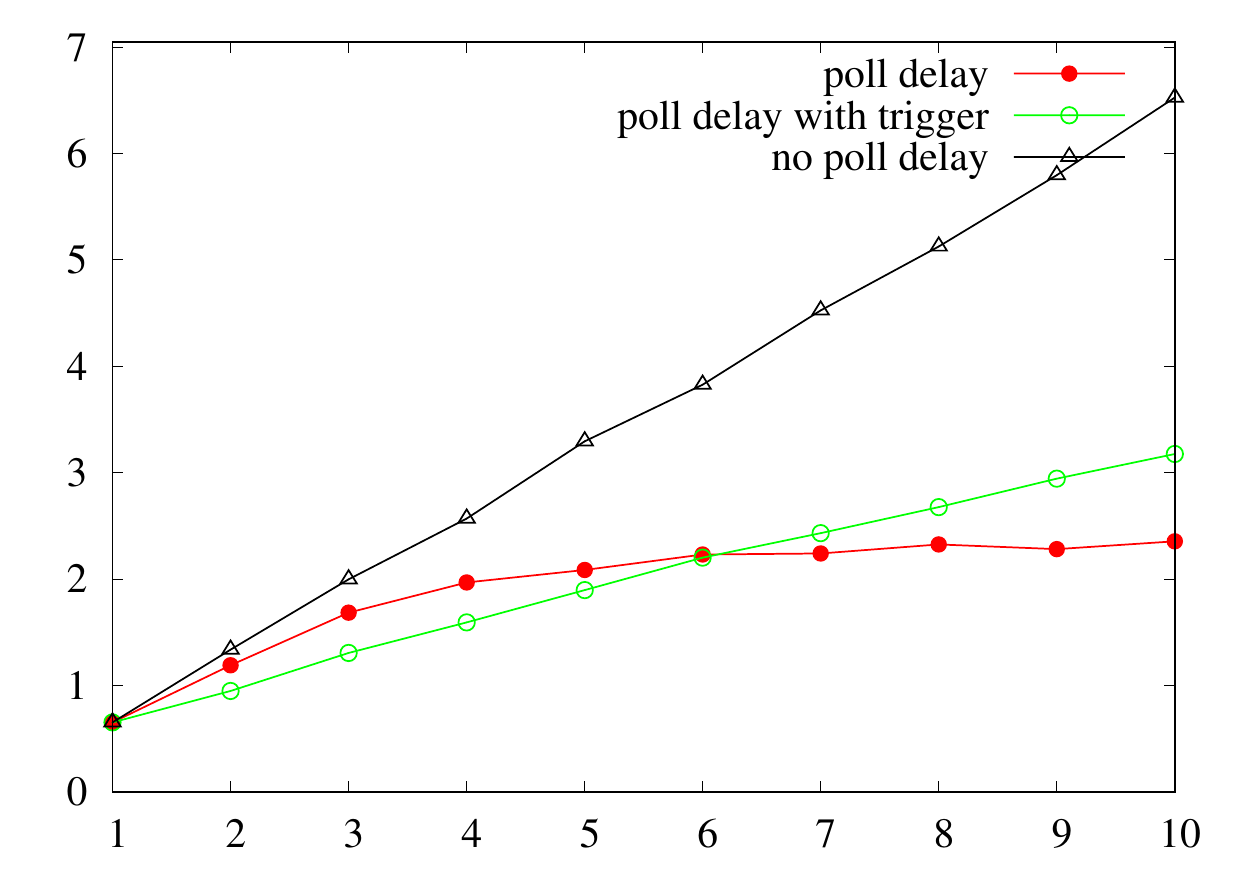}
	\put(-165,55){\rotatebox{90}{$\delay_{0.8}(\ell)$}}
	\put(-80,-5){$\ell$}
	\put(-120,115){Effects of polling delay}
  \caption{ 
%  $\delay_{0.8}(\ell)$ is the 
  %	highest delay $\delay$ that the system can tolerate while achieving a desired  block throughput of  $0.8$.  under $\ell$-\scheme. 
	With a polling delay, % of $\mathsf{Exp}(1/(0.1\delay))$, %$6$-polling 
	% tolerates the slowest network delay while achieving $0.8$ throughput, 
	the performance saturates after $\ell=6$. % and eventually deteriorates at large $\ell$. 
	However, we can continuously harness polling gain 
	if the proposers propose a new block after a fixed time 
	without waiting for all polling to arrive. 
%	the proposers can have a fixed timer 
%	after which 
%	
%	such that after a fixed time the proposer 
%	time before proposing ( fixed wait) performs 
%	better with increasing $\ell$. 
%
	}
	\label{fig:polldelay1}
\endminipage\hfill
\minipage{0.32\textwidth}
   \includegraphics[width=\linewidth]{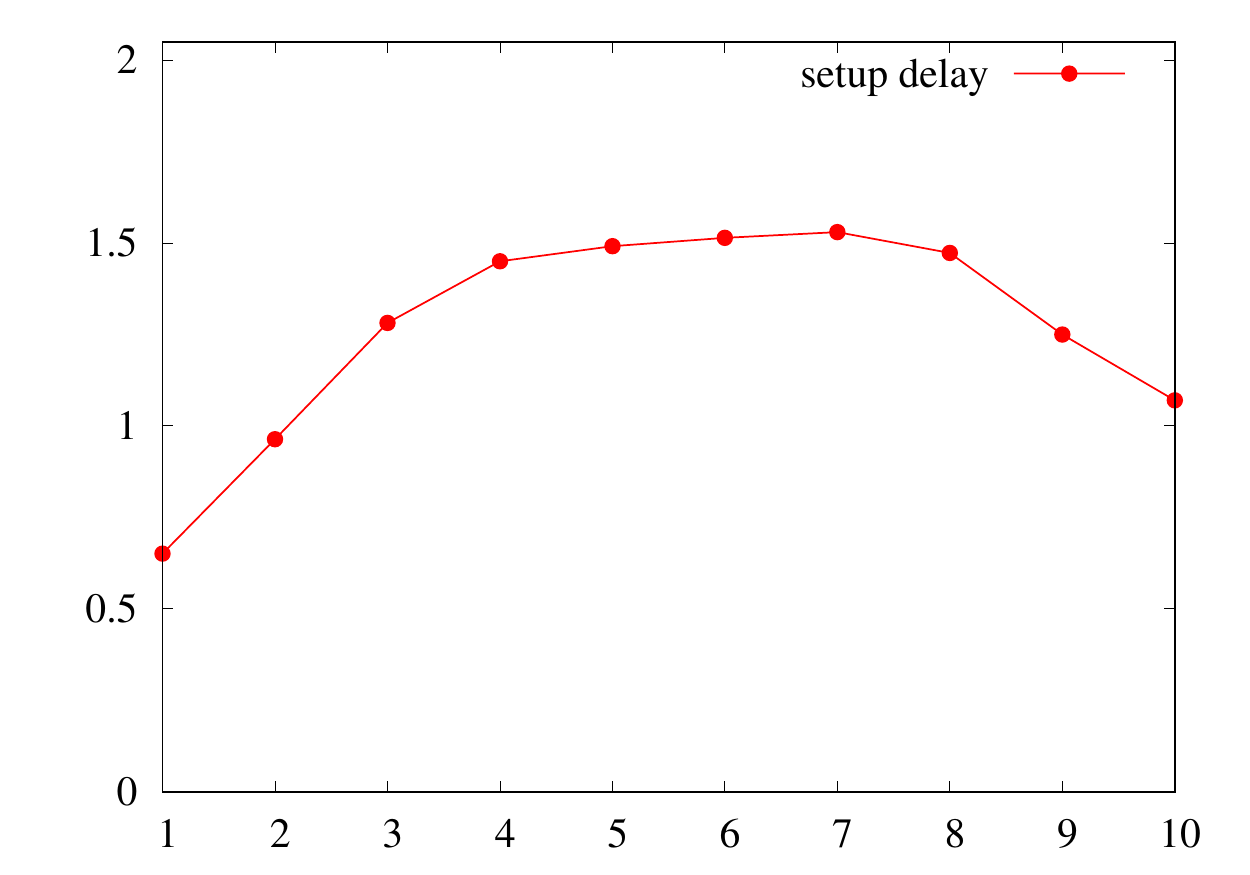}
	\put(-120,115){Effects of setup delay}
  	\put(-165,55){\rotatebox{90}{$\delay_{0.8}(\ell)$}}
	\put(-80,-5){$\ell$}
  \caption{
%  Effects of polling delay. 
%  Average delay $\delay_{0.8}(\ell)$ that achieves desired block throughput  $0.8$, 
%	when $\ell$-\scheme~ is used. 
	We assume a polling delay of $\mathsf{Exp}(1/(0.1\delay))$ but the proposer waits exactly $\tilde{\delay}=0.1\delay$ time before proposing. When there is a setup delay $\propto$ $\ell$, we see an optimal $\ell$, which depends on all system parameters. }
	\label{fig:polldelay2}
\endminipage\hfill
\minipage{0.32\textwidth}%
  \includegraphics[width=\linewidth]{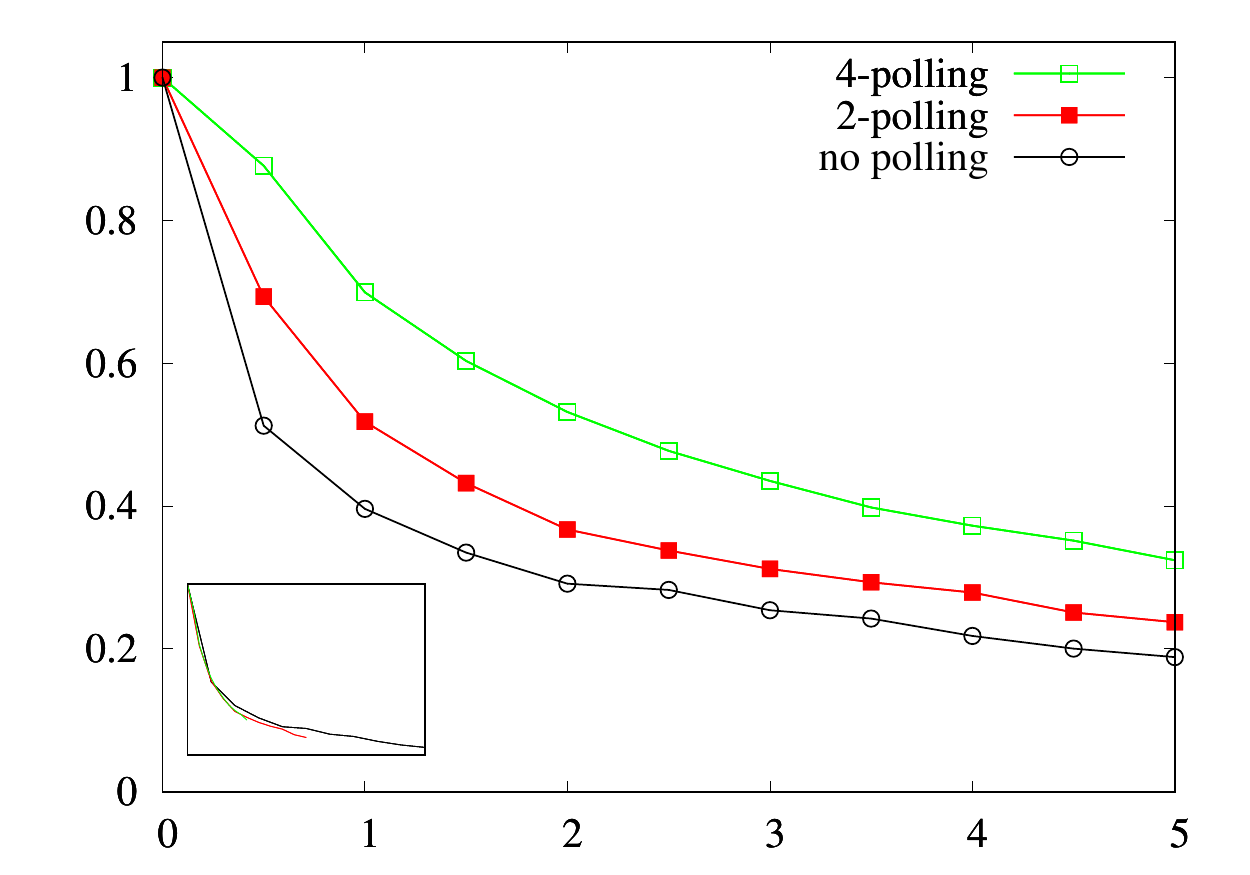}
	\put(-120,115){Heterogeneous networks}
  	\put(-125,-5){mean network delay $\delay$}
	\put(-160,10){\rotatebox{90}{Average block throughput}}
  \caption{Heterogeneous networks enjoy the same polling gain as homogeneous ones.
	Heterogeneous $\ell$-\scheme~ provides a speedup of the network by a factor of about $ \ell$, as shown in the inset 
	where the  x-axis is scaled by $\delay/\ell$. }
	\label{dist2-exponontial}
\endminipage
\end{figure*}

Figure \ref{fig:polldelay1} shows the effect of such polling delay,  
as measured by $\delay_{0.8}(\ell)$, the largest delay $\delay$ that achieves a block throughput of at least 0.8 under $\ell$-\scheme. 
More precisely,
$$
\delay_{0.8}(\ell) = \max \left \{\delay \,:\, \lim_{t\to \infty} \frac{\expect{L(G_t)}}{t} \geq 0.8 \right \}.
$$
Under this model, polling  more nodes means waiting for more responses;
%The gain of polling is less in terms of average throughput, as polling more nodes incurs more polling delays. 
the gains of polling hence saturate for large enough $\ell$, 
and there is an appropriate practical choice of $\ell$ that depends on 
the interplay between the P2P network speed and the polling delay. 

%From the figure, we see that polling performance is a bit degraded compared to zero poll delay scenario, which is expected. However, polling increases performance significantly compared to no polling. Let  us denote the experimental mean of the $L_{chain}(G_{100})$ as $\hat{L}_{chain}(G_{100})$
%We observe that $\hat{L}_{chain}(G_{100}) = 66.58$ for $D=1$, which improves to $\hat{L}_{chain}(G_{100}) = 78.25$ for $2$-polling,
% $\hat{L}_{chain}(G_{100}) = 87.75$ for $3$-polling and $\hat{L}_{chain}(G_{100}) = 95.125$ for $4$-polling. Thus $4$-polling increases the block throughput from by about $43$.
%\purple{Should I write this in terms of complement of block throughput}.

%\begin{figure}[!htb]
%	\centering
%	\includegraphics[width = 0.75\linewidth]{plots/poll_delay1.pdf}
%	\put(-200,75){$\delay_{0.8}(\ell)$}
%	\put(-90,-5){$\ell$}
%	\caption{$\delay_{0.8}(\ell)$ captures the highest delay $\delay$ that achieves a desired  block throughput of  $0.8$ under $\ell$-\scheme. 
%	With a polling delay of $\mathsf{Exp}(1/(0.1\delay))$, %$6$-polling 
%	% tolerates the slowest network delay while achieving $0.8$ throughput, 
%	the performance saturates after $\ell=6$ and eventually deteriorates at large $\ell$. }
%	\label{fig:polldelay1}
%\end{figure}

In practice, there is a strategy to get a large polling gain, even  with delays: 
the proposer polls a large number of nodes, but only waits a fixed amount of time before making a decision. 
Under this protocol, polling more nodes can only help; the only
 cost of polling is the communication  cost. 
The results of our experiments under this protocol are illustrated in Figure \ref{fig:polldelay1} (`poll delay fixed wait' curve).

This implies a gap in our model,  
which does not fully account for the practical cost of polling. 
In order to account for polling costs, we make the model more realistic by assigning a small and constant delay of $0.01\delay$ to set up a connection with a polling node, and assume that the connection setup occurs sequentially for $\ell-1$ nodes. The proposer follows the same strategy as above: waiting for a fixed amount of time before making the decision. We see that under such model, there is a finite optimal $\ell$ as shown in Figure  \ref{fig:polldelay2}.

\begin{figure}[h]
	\includegraphics[width=.6\linewidth]{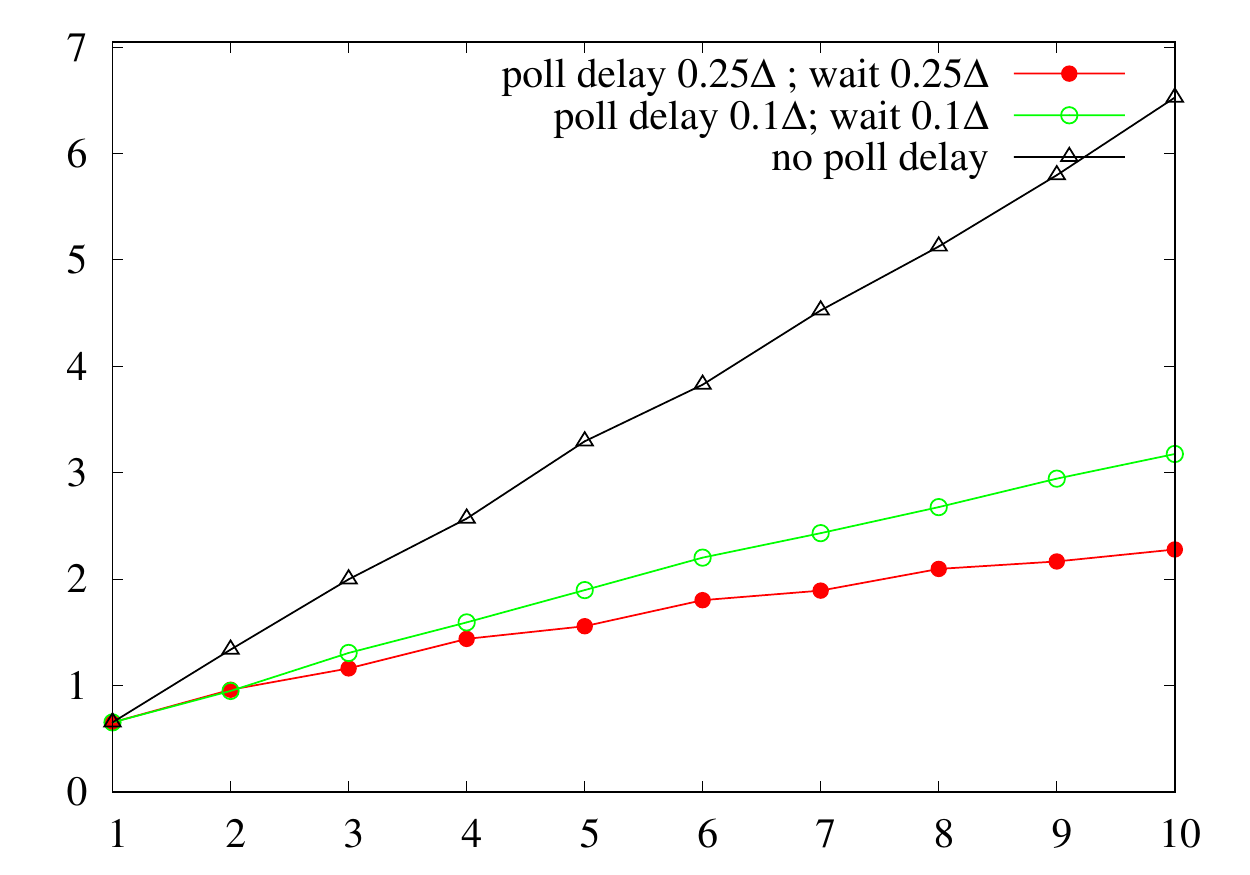}
	\put(-155,40){\rotatebox{90}{$\delay_{0.8}(\ell)$}}
	\put(-70,-5){$\ell$}
	\caption{ 
	The polling gain continues for large $\ell$ even with a more practical choice of a polling delay $0.25\Delta$. 
	}
	\label{fig:morepollingdelay}
\end{figure}

As practical polling delays might be larger than $0.1D$, we compare it to a more practical setting where 
 polling delay is $D/4$ with a threshold wait time of $D/4$ 
in Figure~\ref{fig:morepollingdelay}. 
With this larger delays, the performance is still continuously increasing with $\ell$, and  
%We can see that the performance is still increasing in $\ell$ as expected and 
still provides 250\% improvement at $\ell$=10. 

\subsection{Heterogeneous networks}

The theoretical  and experimental evidence of the benefits of $\ell$-\scheme~ have so far been demonstrated in the context of a homogeneous network: all the nodes in the network have the same bandwidth and processing speeds.  
Further,  individual variation in end-to-end delay due to network traffic is captured by statistically-identical exponential random variables. 
In practice,  heterogeneity is natural: some nodes have stronger network capabilities. 
We model this by 
 clustering the nodes into $h$ different groups 
based on average network speed.  
The speed of a connection is determined by the speed of the slower node.
%The analysis of such network is difficult theoretically, thus we simulate the system to see if polling is effective in such networks.  
We compare the performance of $\ell$-\scheme~ with no polling (which has worse performance and serves as a lower bound). We follow the following uniform polling strategy: Let the delay $\delay$ of a node be a part of the set $\calD = \{ \delay_1,\delay_2,.. ,\delay_h \}$; 
a node's delay is defined as follows: the average delay of transmitting a block across the P2P network  from node with delay $\delay_i$ to a node with delay $\delay_j$ is $\max(\delay_i,\delay_j)$ $\forall i \in [h]$.
%a node's delay refers to the mean delay incurred when that node chooses to broadcast a message.
% then the number, $\ell_i$ of random nodes each node of type $i \in [h]$   polls is the same, i.e.
%  $$ \ell_i = \ell \qquad \forall i \in [h] $$
In Figure~\ref{dist2-exponontial}, we show the performance of a heterogeneous network with $h=2$:  half of the nodes have delay $\delay$ and the others have delay $5\delay$. Every node has the same proposer election probability. 
%With the uniform heterogeneous polling strategy with $\ell_i=\ell$ $\forall i \in [h]$, 
%With $\ell$-polling,
$\ell$-\scheme~ gives a  throughput increase in line Theorem~\ref{thm.dpolling}. 
\subsection{Other practical issues}
\label{sec:security}

There are  remaining three major practical issues.
First, the polling studied in this paper  requires syncing of the complete local blocktree, 
which is redundant and unnecessarily wastes network resources. 
For efficient bandwidth usage, we propose $(\ell,b)$-polling, where 
the polled nodes only send the blocks that were generated between times $t-1$ and $t-b$. 
Secondly,
to ensure timely response from polled nodes, we propose appropriate incentive mechanism, motivated by the reputation systems used in BitTorrent.  
Finally,  a fraction of the participants may deviate from the proposed protocol with  explicit  malicious intent (of causing harm to the key performance metrics). 
It is natural to explore potential security  vulnerabilities exposed by the $\ell$-\scheme~ protocol proposed in this paper. 
All these practical issues are expanded in detail with numerical experiments to support them, in the longer version of this paper available at \cite{longer}. 

%% file: sec6.tex
% !TEX root =  poll.tex
\section{Proofs of the main results}
\label{sec:6}

\subsection{Proof of Theorem~\ref{thm.chainhighprob} }\label{sec:highprobaproof}

We apply Theorem~\ref{thm.dpolling} with general $\ell\geq 1$ and then specialize it to $\ell=1$ to obtain the theorem statement. Denote the chain as $g$, and $e_{j,i,r[1],r[2]}$ as $e_{j,r[1]}$ since here $k = 1$. The event $E_{C,t,g}$ can be written as
\begin{align}
E_{C, t,g} & = \mathds{1}( e_{2,1} = 1) \cdot \mathds{1}(e_{3,1} = 1, e_{3,2} = 1) \cdot \nonumber \\
& \quad \ldots \cdot \mathds{1}(e_{t,1} = 1,e_{t,2} = 1,\ldots,e_{t,t-1} = 1). 
\end{align}

Let $\tilde{E}$ denote the event that every node has proposed or been polled at most once. Conditioned on $\tilde{E}$, and defining $\alpha \triangleq e^{\tilde{\delay}l/\delay}$: 
\begin{align*}
&\mathbb{E}[\mathds{1}(E_{C, t,g})|\tilde{E}]  = \prod_{j = 2}^t \mathbb{E}[\mathds{1}(e_{j,1} = 1,e_{j,2} = 1,\ldots, e_{j,j-1} = 1)|\tilde{E}] \\
& = \prod_{j = 2}^t \prod_{m = 1}^{j-1} (1-e^{\frac{\tilde{\delay}l}{\delay}} e^{-\frac{m \ell}{\delay}})
 = \prod_{j = 1}^{t-1} (1- \alpha e^{-\frac{j \ell} {\delay}})^{t-j}  \leq (1-\alpha e^{-\frac{\ell}{\delay}})^{t-1}. 
\end{align*}
%Denote the arrival time of the information $(\text{Block }j, \text{Point to the Parent of Block }j)$ to node $i$ as $R_{j,i}$. Note that for any $i$ that is not the proposer of block $j$, $R_{j,i} - j$ follows exponential distribution with mean $\delay$. For any $i$ that is the proposer of block $j$, $R_{j,i} = j$. Denote the proposer for block $j$ as $m_j \in [n]$, the event that leads to the final global tree as a chain is 
%\begin{align}\label{eqn.eventwewant}
%E_t & = 
%\mathds{1}(R_{1,m_2} < 2) \mathds{1}(R_{1,m_3}<3, R_{2,m_3}<3) \ldots \mathds{1}(R_{1,m_t}<t, R_{2,m_t}<t,\ldots, R_{t-1, m_t}<t) \\
%& = \prod_{j =1}^t \mathds{1}(E_{j,m_j}). 
%\end{align}
%
%Since we have assumed that $n\gg (\ell t)^2$, it follows from Theorem~\ref{thm.dpolling} that it suffices to compute the expression
%\begin{align}
%\mathbb{E}[\prod_{j = 1}^t \mathds{1}(E_{j,m_j})|\text{all }\{m_j\}_{j =1}^t \text{ are distinct}] 
%\end{align}
%
%Note that if all the $m_j$'s are distinct, we have
%\begin{align}
%& \mathbb{E}[\prod_{j = 1}^t \mathds{1}(E_{j,m_j})|\text{all }\{m_j\}_{j =1}^t \text{ are distinct}]  \\
%& \quad = \prod_{j  =1}^t \mathbb{E}[\mathds{1}(E_{j,m_j})|\text{all }\{m_j\}_{j =1}^t \text{ are distinct}] \\
%& \quad =  \prod_{j = 2}^t \prod_{m = 1}^{j-1} (1-e^{-m/\delay})\\
%&  \quad = \prod_{j = 1}^{t-1} (1-e^{-j/\delay})^{t-j} \\
%& \quad \leq (1-e^{-1/\delay})^{t-1}. 
%\end{align}
%
%Using the notation of Theorem~\ref{thm.dpolling}, we have
%\begin{align}
%F_t(\delay) = \prod_{j = 1}^{t-1} (1-e^{-j/\delay})^{t-j}. 
%\end{align}

We now claim that if $\ell \geq \frac{ \delay \left( \ln t - \ln \ln \frac{1}{\delta} \right)}{1-\tilde{\delay}}$, we have $\mathbb{E}[\mathds{1}(E_{C, t,g})|\tilde{E}]  \geq \delta - o(1)$. Let $c = \ln \frac{1}{\delta}$. 
Indeed, in this case, we have $\alpha e^{-\ell/\delay} \leq {\ln ({1}/{\delta})}/{t}$. Hence, 
\begin{align*}
&\mathbb{E}[\mathds{1}(E_{C, t,g})|\tilde{E}]   \geq \prod_{j = 1}^{t-1} \left( 1 - \frac{c^j}{t^j} \right)^{t-j} \\
&\;\;\;\; = \left( 1- \frac{c}{t} \right)^{t \frac{t-1}{t}} \prod_{j =2}^{t-1} \left( 1 -\frac{c^j}{t^j} \right)^{t-j} 
 \;\stackrel{(a)}\geq \;e^{-c} -o(1) \; =\; \delta - o(1), 
\end{align*}
where $(a)$ follows from Lemma~\prettyref{lmm:limit} and the fact that $\lim_{t \diverge}(1-c/t)^{t-1}=e^{-c}$. Conversely, we show that if $\ell \leq \frac{\delay \left( \ln t - \ln \ln \frac{1}{\delta} \right)}{1-\tilde{\delay}}$, then $\mathbb{E}[\mathds{1}(E_{C, t,g})|\tilde{E}]  \leq \delta +o(1)$. 
Indeed, in this case we have $\alpha e^{-\ell/\delay} \geq {\ln ({1}/{\delta})}/{t}$, and 
\begin{align*}
\mathbb{E}[\mathds{1}(E_{C, t,g})|\tilde{E}] &  \leq (1- c/t)^{t-1} \; =\; e^{-c} +o(1) \; = \;\delta + o(1). 
\end{align*}
\begin{lemma}
\label{lmm:limit}
Let $c>0$ be fixed. Then we have that $\lim_{n \diverge} \sum_{k=2}^{n-1} (n-k) \log(1-{c^k}/{n^k})=0.$
\end{lemma}
%\begin{proof}
%Define $f_n(\cdot): \naturals \to \reals $ as
%$f_n(k) =  (n-k) \log(1-{c^k}/{n^k}) \mathds{1}\{2 \leq k \leq n-1 \}$.
%For each fixed $k \in \naturals$, we have that $\lim_{n \diverge}f_n(k)=0$. Our goal is to show that $\lim_{n \diverge} \int_\naturals f_n(k) d\mu(k)=0$ where $\mu(\cdot)$ is the counting measure on $\naturals$. In view of dominated convergence theorem, hence it suffices to show that there exists a $g: \naturals \to \reals$ such that
%$|f_n(k)| \leq g(k), k \in \naturals \text{ and } \int g d \mu < \infty$.
%Note that for $2 \leq k \leq n-1$,
%\begin{align*}
%|f_n(k)|=  (n-k) |\log(1-\frac{c^k}{n^k})| \stackrel{(a)} \leq (n-k) \frac{{c^k}/{n^k}}{1 - {c^k}/{n^k}} \leq  \frac{nc^k}{n^k-c^k},
%\end{align*}
%where $(a)$ follows from the fact that $|\log(1-x)| \leq \frac{x}{1-x}$ for $x \in [0,1]$. Let $n_0 \in \naturals$ be such that $n_0^k \geq 2c^k$. Hence, for $n \geq n_0$, we have
%$|f_n(k)| \leq \pth{\frac{2}{c}} \pth{\frac{c}{n}}^{k-1} \leq \pth{\frac{2}{c}} \pth{\frac{c}{n_0}}^{k-1} \define g(k)$.
%Clearly $\int_\naturals g(k) d\mu(k)< \infty$. The claim follows. 
%\end{proof}

%\end{lemma}

The proof is included in the extended version \cite{longer}.
Note that the distribution of $\{m_1, m_2,m_3,\ldots,m_t\}$ is independent of $\{R_{ti}:t\geq 1, i\in [n]\}$, hence we could condition on a specific realization of $\{m_2,m_3,\ldots,m_T\}$ and compute the conditional expectation of 
the event that leads to the final global tree as a chain.
%~(\ref{eqn.eventwewant}). 
We  claim: 
\begin{lemma}\label{lemma.individualbound}
For any $T\geq 1$, letting $x = e^{-\lambda}$, we have
$\prod_{j=1}^{T-1} (1-x^j)^{T-j}\leq  \mathbb{E}[E_T|\{m_i: 2\leq i\leq T\}] \leq (1-x)^{T-1}$. 
\end{lemma}

The full proof is included in \cite{longer}. The upper bound uses the independence of the propagation delays, whereas the lower bound relies on the fact that all the $T-1$ events in the $T-1$ indicators functions of $E_T$ are nonnegatively correlated.

Since Lemma~\ref{lemma.individualbound} does not depend on the values of $\{m_i\}_{i = 2}^T$, we know that the bounds apply to $\mathbb{E}[E_T]$ as well. 
For the $d$-polling strategy, it can be verified that both the upper and lower bound computations in Lemma~\ref{lemma.individualbound} are still valid, if we replace $x$ with $x^d$. Indeed, for the upper bound, each block contributes at least $d$ independent random variables; for the lower bound, we can show that the $T-1$ events are positively correlated. 
Now we claim that in order to ensure that $\mathbb{E}[E_T] \geq \delta$, the required number of $d$ is at least approximately
$
d \geq \frac{\ln T - \ln \ln \frac{1}{\delta}}{\lambda}
$
for $T$ large. 
Let $c = \ln \frac{1}{\delta}$. We claim that if $x \leq \frac{c}{T}$, then the probability lower bound is satisfied. Indeed, in this case,
\begin{align*}
\mathbb{E}[E_T] & \geq \prod_{j =1}^{T-1}  \left( 1 - \frac{c^j}{T^j} \right)^{T-j} 
 = \left( 1- \frac{c}{T} \right)^{T \cdot \frac{T-1}{T}} \prod_{j=2}^{T-1} \left( 1- \frac{c^j}{T^j} \right)^{ \frac{T^j}{c^j} \frac{c^j(T-j)}{T^j}} \\
& \geq e^{-c} - o(1) = \delta - o(1)
\end{align*}
as $T\to \infty$.
We also claim that if $x>\frac{c}{T}$, then the probability lower bound is asymptotically not satisfied. Indeed, in this case
\begin{align*}
\mathbb{E}[E_T]  \leq (1-x)^{T-1}  < \left( 1 - \frac{c}{T} \right)^{T-1} 
 = e^{-c} + o(1) = \delta + o(1)
\end{align*} 
as $T\to \infty$. 
Hence, the threshold we aim for should be precisely $x = \frac{c}{T}$. Replacing it with 
$
e^{-\lambda d} = \frac{\ln \frac{1}{\delta}}{T},
$
we get
$
d = \frac{\ln T - \ln \ln \frac{1}{\delta}}{\lambda}. 
$

\subsection{Proof of Theorem \ref{thm.dpolling}}
\label{sec:polling_proof}

\bigskip\noindent
{\bf Part (1).}
 One key observation is that, if every node has only been polled or proposed at most once, i.e,, the set $\{m_{(j,i)}^{(l)}: j\in [t],i \in [k], l\in [\ell]\}$ contains $t k \ell$ distinct nodes, then conditioned on this specific sequence $\{m_{(j,i)}^{(l)}: j\in [t],i \in [k], l\in [\ell]\}$, all the random variables $\{e_{j,i,l,r}: j\in [t], i\in [k],l\in [\ell], r[1]\in [j-1], r[2]\in [k]\}$ are mutually independent. Furthermore, conditioned on this specific sequence, we have
\begin{align}
	\label{eqn.expimplication}
& \mathbb{E}[e_{j,i,l,r}|\{m_{(j,i)}^{(l)}: j\in [t], i\in [k], l\in [\ell]\},  \{\gamma(i)\}_{i =1}^t] \\
& \;\;\;\;  = \;\; 1 - e^{-(\gamma(j)-\gamma(r[1])-\tilde{\delay})/\delay}\;, 
\end{align}
for all $r$ such that $r[1]\in [j-1], r[2]\in [k]$.  
Let $\tilde{E}$ denote the event that $\{m_{(j,i)}^{(l)}:j\in [t],i\in [k],l\in [\ell]\}\text{ are distinct}$. It follows from the definition of local attachment protocol $\mathcal{C}$ that 
$\mathbb{E}[e_{j,i,l,r}|\tilde{E}, , \{\gamma(i)\}_{i =1}^t] = 1 - e^{-(\gamma(j)-\gamma(r[1])-\tilde{\delay})/\delay}$
for all $r$ such that $r[1]\in [j-1], r[2]\in [k]$.  
Note that the event $E_{C,t,g} = \{G_t = g\}$ only depends on $\{m_{(j,i)}^{(l)}:j\in [t],i\in [k],l\in [\ell]\}$ and $\{e_{j,i,r}: j\in [t],i \in [k], r[1]\in [j-1], r[2]\in [k]\}$ plus some additional outside randomness. Since 
$
e_{j,i,r} = 1 \Leftrightarrow \sum_{l\in [\ell]} e_{j,i,l,r} \geq 1,
$
it follows from the independence of $e_{j,i,l,r}$ and equation~(\ref{eqn.expimplication}) that 
\begin{align} \label{eqn.jir}
\mathbb{E}[e_{j,i,r}|\tilde{E}, \{\gamma(i)\}_{i =1}^t] = 1 - e^{-(\gamma(j)-\gamma(r[1])-\tilde{\delay})\ell/\delay}
\end{align}
all $r$ such that $r[1]\in [j-1], r[2]\in [k]$.  
Hence, we have
\begin{align*}
\mathbb{P}(\tilde{G}_t = g) & = \mathbb{E}\left[ \mathds{1}(E_{C,t,g})|\tilde{E}, \{\gamma(i)\}_{i =1}^t \right]  = F(\{\gamma(i)\}_{i =1}^t, \frac{\delay}{\ell}, \tilde{\delay}, g, \mathcal{C}). 
\end{align*}
Now we show the second part of Theorem~\ref{thm.dpolling}. Denote by $A = \{g_1,g_2,\ldots,g_A\}$ any collection of distinct tree structures that $G_t$ may take values in. Then, we have
\begin{align}
& \mathbb{P}(G_t \in A| \{\gamma(i)\}_{i =1}^t)  = \mathbb{E}\left[ \sum_{i = 1}^A \mathds{1}(E_{C,t,g_i}) \Bigg | \{\gamma(i)\}_{i =1}^t \right] \\
%& = \mathbb{P}(\tilde{E}|\{\gamma(i)\}_{i =1}^t)  \mathbb{E}[\sum_{i = 1}^A \mathds{1}(E_{C,t,g_i}) |\tilde{E},\{\gamma(i)\}_{i =1}^t] \nonumber \\
%& \quad +(1- \mathbb{P}(\tilde{E}|\{\gamma(i)\}_{i =1}^t) ) \mathbb{E}[\sum_{i = 1}^A \mathds{1}(E_{C,t,g_i}) | \tilde{E}^c,\{\gamma(i)\}_{i =1}^t] \\
%& =  \mathbb{E}[\sum_{i = 1}^A \mathds{1}(E_{C,t,g_i}) |\tilde{E},\{\gamma(i)\}_{i =1}^t] + (1- \mathbb{P}(\tilde{E}|\{\gamma(i)\}_{i =1}^t) )\nonumber \\
%& \quad \times \left( \mathbb{E}[\sum_{i = 1}^A \mathds{1}(E_{C,t,g_i}) | \tilde{E}^c,\{\gamma(i)\}_{i =1}^t] - \mathbb{E}[\sum_{i = 1}^A \mathds{1}(E_{C,t,g_i}) |\tilde{E},\{\gamma(i)\}_{i =1}^t] \right) \\
& = \mathbb{P}(\tilde{G}_t \in A)  + (1- \mathbb{P}(\tilde{E}|\{\gamma(i)\}_{i =1}^t) ) \nonumber \\
& \quad \times \left( \mathbb{E}[\sum_{i = 1}^A \mathds{1}(E_{C,t,g_i}) | \tilde{E}^c,\{\gamma(i)\}_{i =1}^t] - \mathbb{E}[\sum_{i = 1}^A \mathds{1}(E_{C,t,g_i}) |\tilde{E},\{\gamma(i)\}_{i =1}^t] \right). 
\end{align}

It follows from the birthday paradox compution~\cite[Pg. 92]{mitzenmacher2005probability} that 
$1- \mathbb{P}( \tilde{E}|\{\gamma(i)\}_{i =1}^t) \leq {kt\ell(kt\ell-1)}/{2n}$.
Hence, we have shown that for any measurable set $A$ that $G_t$ or $\tilde{G}_t$ take values in, we have
$| \mathbb{P}(G_t \in A|\{\gamma(i)\}_{i =1}^t) - \mathbb{P}(\tilde{G}_t \in A) |  \leq 1-\mathbb{P}(\tilde{E}|\{\gamma(i)\}_{i =1}^t) 
 \leq \frac{kt\ell(kt\ell-1)}{2n}$. 
The result follows from the definition of the total variation distance
$\text{TV}(P_{G_t|\{\gamma(i)\}_{i =1}^t}, P_{\tilde{G}_t}) = \sup_{A} | \mathbb{P}(G_t \in A|\{\gamma(i)\}_{i =1}^t) - \mathbb{P}(\tilde{G}_t \in A) |$. 

\noindent
{\bf Part (2).}
We note that there exists some function $L(\{\gamma(i)\}_{i =1}^t, \frac{\delay}{\ell}, \tilde{\delay},\mathcal{C})$ independent of all the parameters in the model such that the expectation of the longest chain of $\tilde{G}_t$ is equal to $L(\{\gamma(i)\}_{i =1}^t, \frac{\delay}{\ell}, \tilde{\delay},\mathcal{C})$. To obtain the final result, it suffices to use the variational representation of total variation distance, $\text{TV}(P,Q) = \sup_{f: |f|\leq \frac{1}{2}} \mathbb{E}_P f - E_Q f$,
and taking $f = \frac{1}{t}\cdot \left( L_{\text{Chain}}(G_t) - t/2 \right)$, upon noticing that the length of the longest chain in the tree $G_t$ is at most $t$.

%% file: related.tex
\section{Related Work}
\label{sec:related}

Four main approaches exist for reducing forking. 
%The first is to reduce proposer diversity. 
%The second is to use forking to enhance throughput. 
%The third is to algorithmically resolve forking before moving to the next block.

\emph{(1) Reducing proposer diversity.}
%Forking is caused by the delay associated with the most recently-proposed block reaching the next proposer(s). 
A natural approach is to make the same node propose consecutive blocks;
for instance, Bitcoin-NG \cite{bitcoin-ng}
proposers use the longest-chain fork choice rule, but within a given time epoch, only a single proposer can propose blocks. 
This allows the proposer to quickly produce blocks without forking effects. 
Although Bitcoin-NG has high throughput, it exhibits a  few problems. 
When a single node is in charge of block proposal for an extended period of time, attackers may be able to learn that node's IP address and take it down. 
%Second, it suffers from high confirmation delay. To confirm a transaction in a longest-chain protocol, we require a threshold number of independently-selected proposers to append blocks to the chain containing that transaction. Since Bitcoin-NG elects a new proposer only once every epoch, this takes time comparable to Nakamoto consensus.
The idea of  fixing  the proposer is also used in other protocols, such as Thunderella \cite{thunderella} and ByzCoin \cite{byzcoin}. 
%which are also vulnerable to attacks on the proposer's IP address.

\emph{(2) Embracing forking.} Other protocols  use forking to contribute to throughput. 
Examples include GHOST \cite{ghost}, PHANTOM \cite{phantom}, SPECTRE \cite{spectre}, and Inclusive/Conflux \cite{inclusive,conflux}.
GHOST describes a fork choice rule that tolerates honest forking by building on the heaviest subtree of the blocktree. %; it is described more carefully in Section \ref{sec:2b}.
SPECTRE, PHANTOM, and Conflux instead use existing fork choice rules, but build a directed acyclic graph (DAG) over the produced blocks to define a transaction ordering. 
A formal understanding of such DAG-based protocols is  evolving; their security properties are not yet well-understood. 

\emph{(3) Structured DAGs.}
A related approach is to allow \emph{structured} forking. 
The Prism consensus mechanism explicitly co-designs a consensus protocol and fork choice rule to securely deal with concurrent blocks, thereby achieving optimal throughput and latency \cite{prism}.
The key intuition is to run many concurrent blocktrees, where a single proposer tree is in charge of ordering transactions, and the remaining voter trees are in charge of confirming blocks in the proposer tree. 
%Our approach differs from \cite{prism} in that 
\scheme~ is designed to be integrated into \emph{existing} consensus protocols, whereas \cite{prism} is a new consensus protocol.
%Indeed, since each of the blocktrees in Prism uses the longest-chain fork choice rule, \scheme~ can be used to reducing forking in each individual Prism blocktree.

\emph{(4) Fork-free consensus.}
Consensus protocols like Algorand \cite{algorand2}, Ripple, and Stellar \cite{stellar} prevent forking entirely by conducting a full round of consensus for every block. 
%Disagreements about the next block  are resolved immediately. 
Although  voting-based consensus protocols consume additional time for each block, they may improve overall efficiency by removing the need to resolve forks later; this hypothesis remains untested.
A  challenge in such protocols is that BFT voting protocols can be communication-intensive, and require a known set of participants. 
Although some work addresses these challenges \cite{byzcoin,hybrid},  many industrial blockchain systems running on BFT voting protocols require some centralization. 
%This approach reflects a fundamental question regarding consensus in blockchains: is forking  

{Our approach}
%\scheme~  complements   prior approaches by providing a networking protocol that is compatible with existing consensus algorithms. 
%Rather than completely eliminating forking, we run a round of polling to improve the proposer's blocktree estimate.  
 can be viewed as a partial execution of a polling-based consensus protocol.
Polling has long been used  in consensus protocols \cite{cruise2014probabilistic,abdullah2015global,fischer1985textordfeminineimpossibility,avalanche}.
Our approach differs in part because we do not use polling to reach complete consensus, but  to reduce the number of  inputs to a (separate) consensus protocol. 
%Hence, \scheme~ can be used with many consensus protocols, especially those proposed for proof-of-stake (PoS) systems.
%; we discuss these applications in Section~\ref{sec:5}. 
%\purple{Our protocol requires leader election mechanism which is independent of the parent block(as the parent block might change after polling), so it might not be directly useful for Bitcoin or Ethereum)}
%% I think it is fine, as the idea of polling could be applied in other scenarios. For example, in bitcoin one could poll  before starting mining to update the current block tree. As long as there is forking, polling idea could be applied. (Sewoong)
%For example, it can naturally be integrated with approaches (1) and (2). 
%For approach (1), polling could occur between epochs to enfor

%% file: conclusion.tex
%\vspace{-0.5cm}
\section{Conclusion}
\label{sec:7}

In this paper, we propose $\ell$-polling as a technique for improving block throughput in proof-of-stake cryptocurrencies. 
We show that for small $\ell$, $\ell$-polling has the same effect on block throughput as if the mean network delay were reduced by a factor of $\ell$.
This simple, lightweight method improves throughput without substantially altering either the underlying consensus protocol or the  network.
Several open questions remain, particularly with regards to analyzing adversarial behavior in $\ell$-polling. We have avoided them in this paper by proposing a symmetric version of the protocol (cf.\ Section~\ref{sec:security}), but even within the original $\ell$-polling protocol, it is unclear how much an adversary could affect block throughput and/or chain quality by responding untruthfully to poll requests. 
%Preliminary results suggest that this effect is small. %; this is an active area of research. 